\documentclass[useAMS,usenatbib]{mn2e}
\usepackage{graphicx}
\usepackage{subfigure}

\title[Towards Precise Ages and Masses of Free Floating Planetary Mass Brown Dwarfs]{Towards Precise Ages and Masses of Free Floating Planetary Mass Brown Dwarfs}
\author[J. I. Canty et al.]
{J. I. Canty$^{1}$\thanks{E-mail:j.canty2@herts.ac.uk}
P.W. Lucas$^{1}$ P.F. Roche$^{2}$ D.J. Pinfield$^{1}$\\
$^{1}$Centre for Astrophysics Research, University of Hertfordshire, College Lane, Hatfield, AL10 9AB, UK\\
$^{2}$Astrophysics Department, University of Oxford, 1 Keble Road, Oxford OX1 3RH}
\begin{document}

\pagerange{\pageref{firstpage}--\pageref{lastpage}} \pubyear{2013}

\maketitle

\label{firstpage}

\begin{abstract}
Measurement of the substellar initial mass function (IMF) in very young clusters is hampered by the possibility of the age spread of cluster members. This is particularly serious for candidate planetary mass objects (PMOs), which have a very similar location to older and more massive brown dwarfs on the Hertzsprung-Russell Diagram (HRD). This degeneracy can be lifted by the measurement of gravity sensitive spectral features. To this end we have obtained medium resolution (R$\approx$5000) Near-infrared Integral Field Spectrometer (NIFS) K~band spectra of a sample of late~M-/early~L-type dwarfs. The sample comprises old field dwarfs and very young brown dwarfs in the Taurus association and in the $\sigma$~Orionis cluster. We demonstrate a positive correlation between the strengths of the 2.21~$\mu$m NaI doublet and the objects' ages. We demonstrate a further correlation between these objects' ages and the shape of their K~band spectra. We have quantified this correlation in the form of a new index, the H$_{2}$(K) index. This index appears to be more gravity-sensitive than the NaI doublet and has the advantage that it can be computed for spectra where gravity-sensitive spectral lines are unresolved, while it is also more sensitive to surface gravity at very young ages ($<$10~Myr) than the triangular H~band peak. Both correlations differentiate young objects from field dwarfs, while the H$_{2}$(K) index can distinguish, at least statistically, populations of $\sim$1~Myr objects from populations of $\sim$10~Myr objects. We applied the H$_{2}$(K) index to NIFS data for one Orion nebula cluster (ONC) PMO and to previously published low resolution spectra for several other ONC PMOs where the 2.21~$\mu$m NaI doublet was unresolved and concluded that the average age of the PMOs is $\sim$1~Myr.
\end{abstract}


\section{Introduction}
Deriving the substellar IMF requires a representative sample of objects whose ages are well-constrained. Evolutionary models can then be used to assign masses and radii to these objects. However, because cooling is more rapid for less massive bodies, brown dwarfs of any given effective temperature have a well known degeneracy between age and mass. In very young (1-10~Myr) brown dwarfs this problem is most serious for the coolest and least massive objects with spectral types $>$ M8 and masses $<$ 25~M$_{\rm{Jup}}$ (see, for example, Figure 5 of~\citealt{weights09}, hereafter W09), where objects with planetary masses and ages of $\sim$1~Myr are not well separated in the HRD from brown dwarfs with ages of $\sim$10~Myr and masses 3 times higher. The Lyon DUSTY isochrones~\citep{allard01} and those of~\citet{d'antona97} agree in this. In order to derive the substellar IMF, this age-mass degeneracy must be lifted.

It has been shown that source ages can be measured using gravity sensitive alkali metal absorption lines in the optical or near-infrared (\citealt{martin96};~\citealt{steele95};~\citealt{gorlova03};~\citealt{allers07}, hereafter A07;~\citealt{riddick07};~\citealt{luhman07};~\citealt{close07}). This technique can differentiate 1 Myr objects from 5 Myr objects, and 5-10 Myr objects from field dwarfs, since very young substellar objects rapidly contract to smaller radii. The radii of evolved field brown dwarfs have little dependence on age or mass~\citep{burrows93}, while the radii of very young brown dwarfs can be up to six times the size of their final equilibrium radii~\citep{stassun06}. In consequence, their surface gravities ($g=GM/R^2$) can be substantially lower than those of older, more massive dwarfs of the same spectral type.

Theoretical models show that the luminosities of brown dwarfs decrease rapidly over time as they radiate away the internal energy supplied by gravity during the formation process (e.g.~\citealt{burrows01}, hereafter B01). Therefore, many analyses have focussed on young clusters, where the higher luminosities of brown dwarfs at young ages allows the mass function to be derived for masses as low as 5-15M$_{\rm{Jup}}$.

Among the uncertainties in these models are those dealing with the ages of the objects and the accuracy of the models themselves. These uncertainties are most significant in the cases of very young brown dwarfs in pre-main sequence clusters. 

One uncertainty concerns a brown dwarf's initial energy. A `hot start' brown dwarf forms from the collapse of a molecular cloud. A brown dwarf formed in a `cold start' via core accretion has less gravitational potential energy stored as internal energy so that, mass for mass, the brown dwarf is cooler and less luminous than a `hot start' brown dwarf. After $\sim$2 Myr, runaway gas accretion produces a peak in the luminosity of `cold start' brown dwarfs~\citep{marley07}. For the next $\sim$1 Myr and regardless of their mass, `cold start' brown dwarfs outshine their `hot start' counterparts, after which they cool quickly and become less luminous than `hot start' brown dwarfs of a similar mass. This implies that `cold start' brown dwarfs older than a few Myr would need to have higher masses in order to account for their observed luminosities. We note that since `cold start' brown dwarfs form from a protoplanetary disk, this formation model is unlikely to apply to isolated brown dwarfs.

The constant mass accretion rate described in the standard model of star formation (~\citealt{shu77};~\citealt{terebey84}) may give rise to the ``luminosity problem", exemplified in the case of young nearby T Tauri stars with solar luminosities. These stars are under-powered in the sense that they should have several solar luminosities if they are descending approximately vertical Hayashi tracks on the HRD~\citep{kenyon90}, depending on how young they are when they are observed. Time-variable accretion may resolve the ``luminosity problem". It has been shown that episodic bouts of accretion rather than a spread of ages can also explain the luminosity spread in young clusters~(\citet{baraffe12}, hereafter B12).

These various uncertainties become less of a problem after only a few Myr, but they must be considered when dealing with the youngest brown dwarfs.

In this paper we investigate spectral signatures of age for substellar objects in pre-main sequence clusters. Previous work by this group and others has established the existence of a population of PMOs in the ONC (\citealt{lucas06}, hereafter L06;~\citealt{riddick07}). A statistical analysis of the luminosity function by W09 indicates that most substellar objects in the cluster have ages of order 1~Myr, and it is unlikely that the PMO candidates represent a tail of older and more massive objects. However, the size of the planetary mass population cannot be tightly constrained until their ages are known with better precision. 

The observations described in this paper were intended to produce the highest resolution spectra yet obtained (R$\approx$5000) of three PMOs in the ONC and a sample of brighter calibrator brown dwarfs with known ages and/or gravities, including three objects on the deuterium-burning threshold in the $\sigma$~Orionis cluster. In the event, we were unable to obtain useful data for two of the three ONC targets. However, the high quality of the data obtained for all of the calibrators has enabled a useful investigation of the effects of surface gravity on the K~band spectra of very young brown dwarfs.

Our aim was to use medium resolution spectroscopy to investigate the sensitivity of the NaI doublet at 2.21$\mu$m to surface gravity, and hence age, while using the CaI line at 2.26$\mu$m and the CO absorption bands at 2.29$\mu$m and 2.32$\mu$m to correct for any metallicity variations or veiling by hot dust. It was also intended to use these spectra to provide more precise spectral types than had been possible previously, and also to confirm that the IMF extends below 10M$_{\rm{Jup}}$. An unexpected by-product of the investigation was the discovery that the slope of the K~band also has strong sensitivity to gravity, which appears to provide a useful age indicator. While the effect of surface gravity on the K~band spectra of pre-main sequence brown dwarfs has previously been noted (\citealt{luhman04a}; A07), it has not been studied in detail. 

\section{Structure of this Paper}
Section 3 of this paper describes how our sample of brown dwarf calibrators and PMOs were chosen.

Section 4 describes how these objects were observed, and how their spectra were extracted.

Results are reported in Section 5. Section 5.1 contains the objects' K~band spectra. Section 5.2 describes how the strengths of the NaI lines and other spectral features were determined. Section 5.3 deals with the objects' revised spectral types.

Section 6 relates our findings regarding collisionally induced absorption by H$_2$ in the K~bandpass. Section 6.1 describes the derivation of a new index, the H$_{2}$(K) index, to measure this absorption and how the index is related to the objects' ages. Section 6.2 uses models to show changes in the K~band slope with surface gravity. Section 6.3 tests the H$_{2}$(K) index for the effects of hot dust, extinction, noise and metallicity, and examines whether the H$_{2}$(K) index varies as a function of spectral type.

Section 7 describes how the H$_{2}$(K) index was examined using an extended dataset of spectra from the literature.

Our results are discussed in Section 8. Section 8.1 contains a more detailed discussion of the results for our sample of brown dwarf calibrators. Section 8.2 discusses our observed spectrum of an ONC PMO. Section 8.3 describes a distinctive water absorption pattern in the K~band which may be an additional diagnostic in classifying spectra. 

Section 9 compares the strength of neutral alkali metal lines with the H$_{2}$(K) index as age indicators. 

Section 10 contains our conclusions.

\section{Selection of Targets}
In the last few years, large populations of brown dwarfs have been discovered in several very young clusters where the mass function appears to extend below the deuterium-burning threshold of 0.012-0.013M$_{\odot}$ (\citealt{lucas00};~\citealt{zapatero00};~\citealt{scholz11};~\citealt{scholz12}). The ONC contains the largest known sample of very young brown dwarfs, and therefore is an ideal site to obtain good statistics on the IMF at planetary masses and to gain insights into low-mass brown dwarf formation processes. The formation mechanism responsible for the low masses of the substellar objects is currently unknown.

The three ONC PMOs are 183$-$729 (18), 152$-$717 (27) and 137$-$532 (172) (\citealt{lucas00};~\citealt{lucas05}, hereafter LRT05) (the numbers in brackets are the source catalogue numbers quoted in LRT05). They were selected because each has a bright star $I_{MAG}$=15$-$16 and $R_{MAG}$=17$-$18 within $\sim$25$\arcsec$ that could potentially be used as a tip/tilt guide star for laser-guided adaptive optics (AO) and they are bright enough for medium resolution ground based spectroscopy. We note that there is only a modest range of apparent K magnitudes amongst the spectroscopically confirmed PMOs in L06. Their PMO status is based on their spectral types of $\sim$M9$-$L0, their low surface gravities (L06; W09), and their low luminosities. Low gravity was determined by comparing these objects' H and K~band pseudo-continuum profiles with those of other low gravity brown dwarfs and high gravity field dwarfs in low resolution spectra. These features confirmed that these objects were cluster members, and therefore that their ages are unlikely to be more than $\sim$10~Myr. The PMOs are within a few arcminutes of the centre of the ONC, so they are likely to be very young ($\sim$1~Myr) objects, rather than members of the more dispersed and slightly older population that extends over a few degrees on the sky.

In the event, useful observations were obtained for only one ONC PMO, 152$-$717. This source has apparent magnitude K=17.6, which is typical of the sample in L06. It was not possible to obtain a guide star lock for the other two PMOs, owing to the bright background of the nebulosity. We note that this observing method could be more successfully employed in future when the patrol radius of the Gemini Altair AO system has been increased, allowing the use of brighter guide stars.

The calibrators consisted of two field dwarfs, 2MASS~0345+25, hereafter 2MASS~0345~\citep{kirkpatrick97}, BRI~0021$-$0214, hereafter BRI~0021~\citep{irwin91}, three 3-7~Myr objects, $\sigma$~Orionis~51~\citep{zapatero00}, $\sigma$~Orionis~71~\citep{barrado02}, $\sigma$~Orionis J053849.5$-$024934, hereafter $\sigma$~Orionis J053$-$024, and five 1-2~Myr objects, KPNO-Tau~1~\citep{briceno02}, KPNO-Tau~4~\citep{briceno02}, KPNO-Tau~12~\citep{luhman03}, 0457$+$3015, and 2MASS~0535$-$0546~\citep{stassun06}. 

The calibrators were all chosen because they have similar spectral types to the ONC PMOs, with the exception of the eclipsing binary 2MASS~0535$-$0546 (type M7) which was felt to be worthy of observation owing to the precisely known surface gravities of the components.

The three $\sigma$~Orionis objects were selected because there was fairly good prior evidence that they were bona fide cluster members, as opposed to contaminating field dwarfs in front of this rather diffuse cluster. This evidence was either from previous spectroscopy of alkali metal lines ($\sigma$~Orionis~51, \citep{mcgovern04}) or from detection of mid-infrared excess and/or spectroscopic evidence ($\sigma$~Orionis~71, $\sigma$~Orionis~J053$-$024, \citep{caballero07}). Being older, they can be used to compare objects of slightly greater mass and gravity. 

Further selection criteria for the 1-2~Myr calibrators in the Taurus-Auriga molecular cloud complex and field dwarf calibrators were that they are bright objects with little or no extinction and that previous studies have found them to be fairly typical objects with no obvious veiling by a circumstellar accretion disc (see above references). While the 1-2~Myr calibrators are more massive than the ONC PMOs, the Lyon DUSTY isochrones predict that they also have larger radii, such that their surface gravities are very similar.

\begin{figure*}
   \centering
   \begin{subfigure}
         \centering
          \includegraphics[height=4.35in, width=2.in]{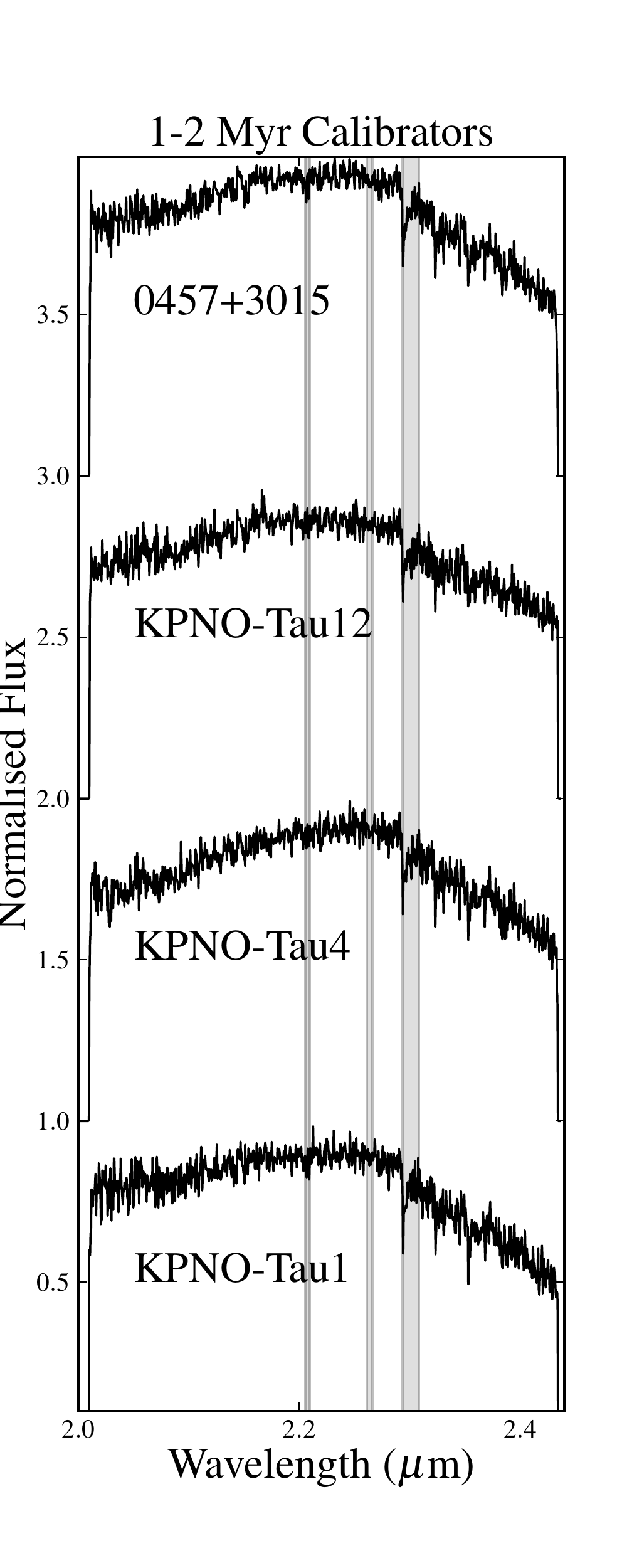}
   \end{subfigure}
   \begin{subfigure}
         \centering
          \includegraphics[height=4.35in, width=2.in]{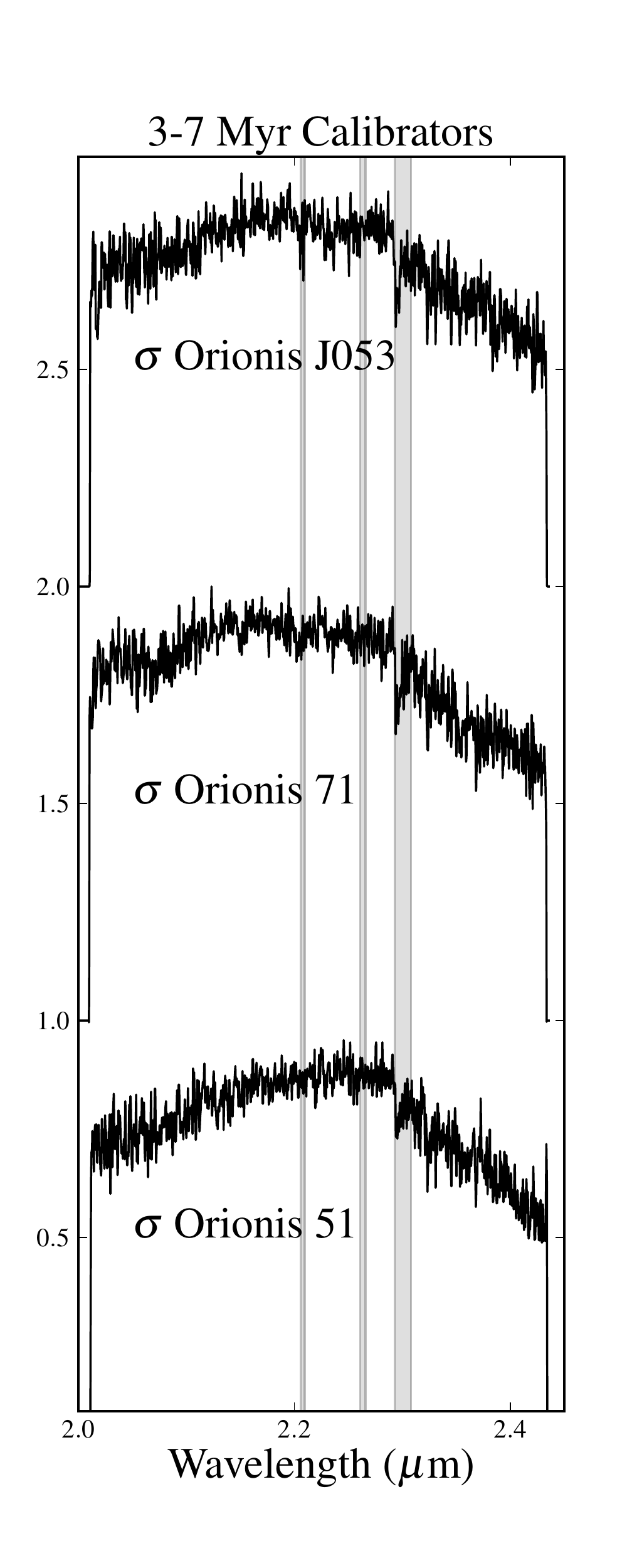}
   \end{subfigure}
   \begin{subfigure}
         \centering
          \includegraphics[height=4.35in, width=2.in]{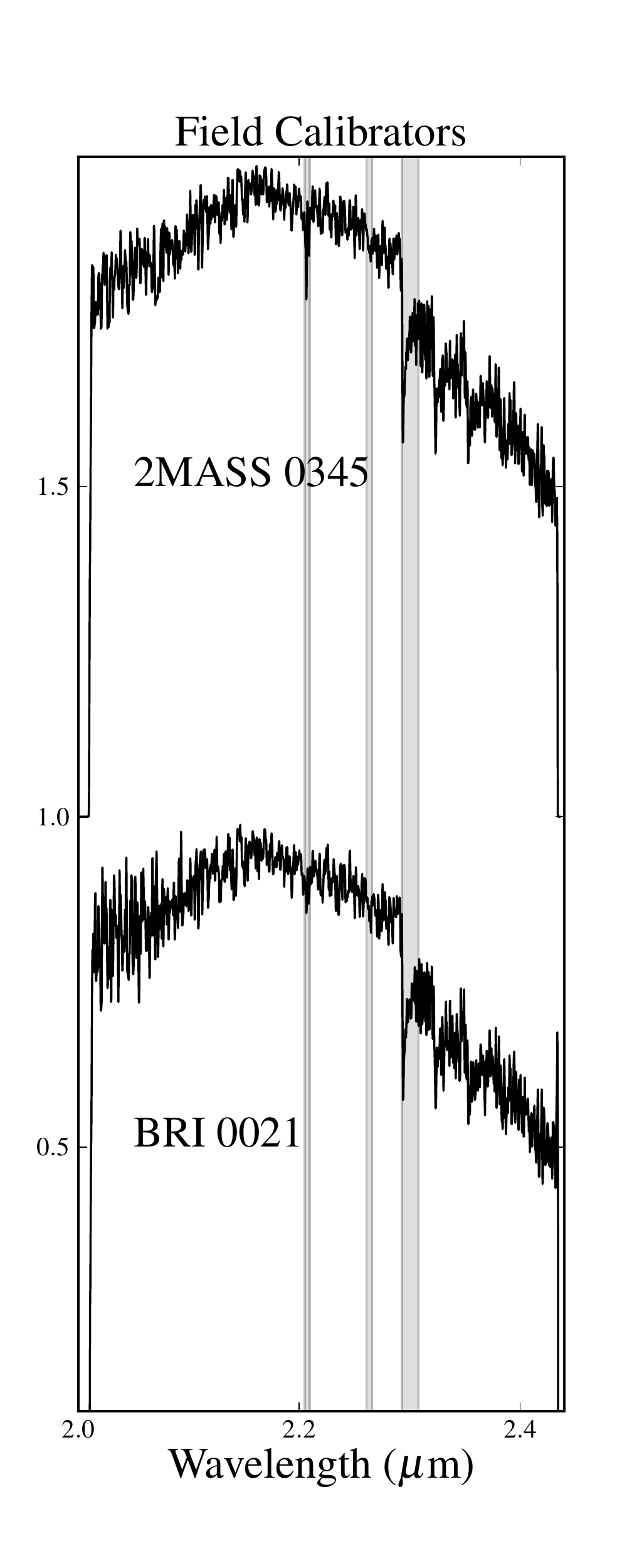}
   \end{subfigure}  
   \centering
   \begin{subfigure}
         \centering
          \includegraphics[height=4.35in, width=2.in]{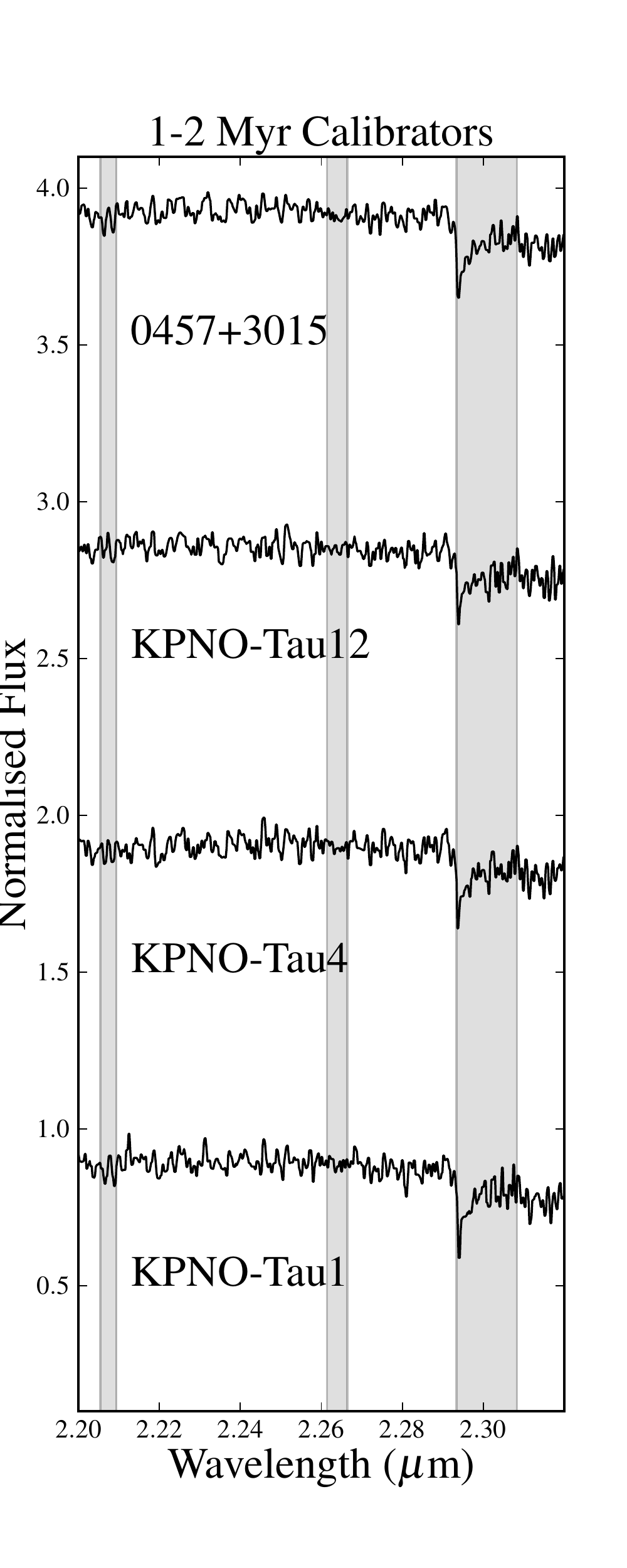}
   \end{subfigure}
   \begin{subfigure}
         \centering
          \includegraphics[height=4.35in, width=2.in]{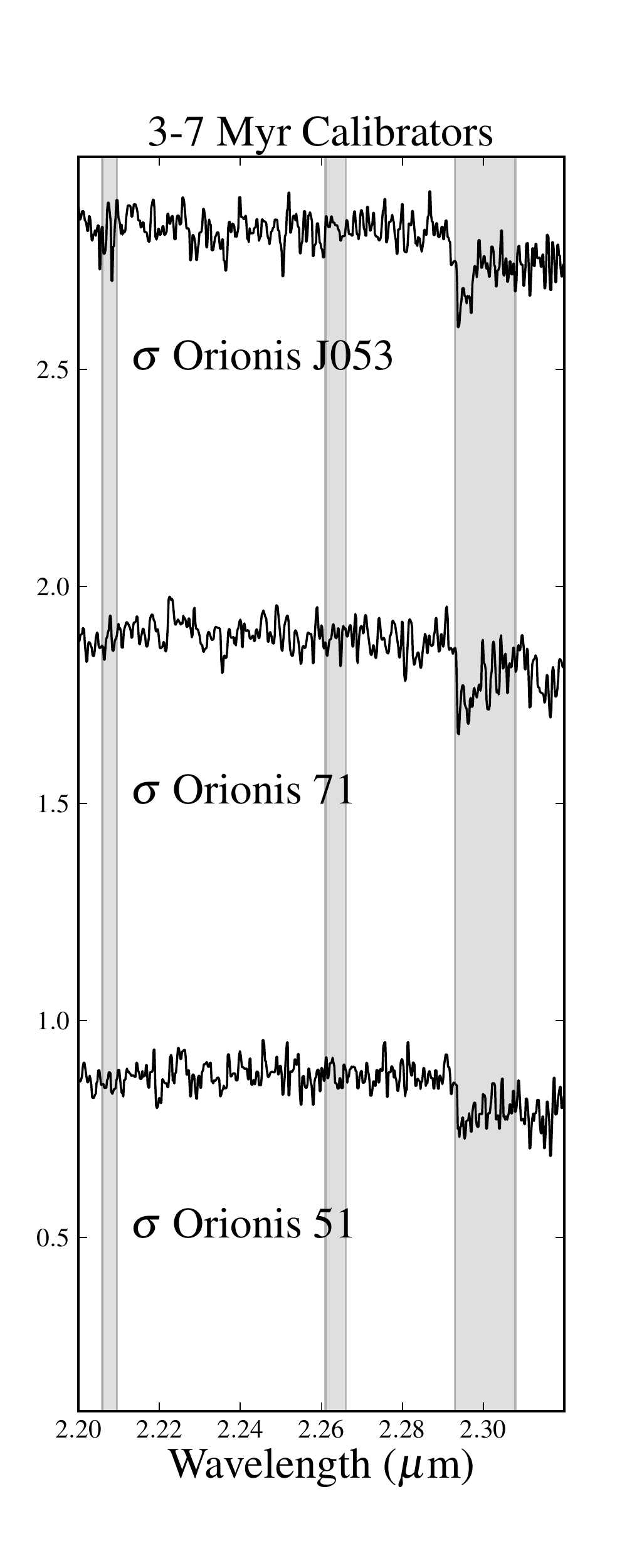}
   \end{subfigure}
   \begin{subfigure}
         \centering
          \includegraphics[height=4.35in, width=2.in]{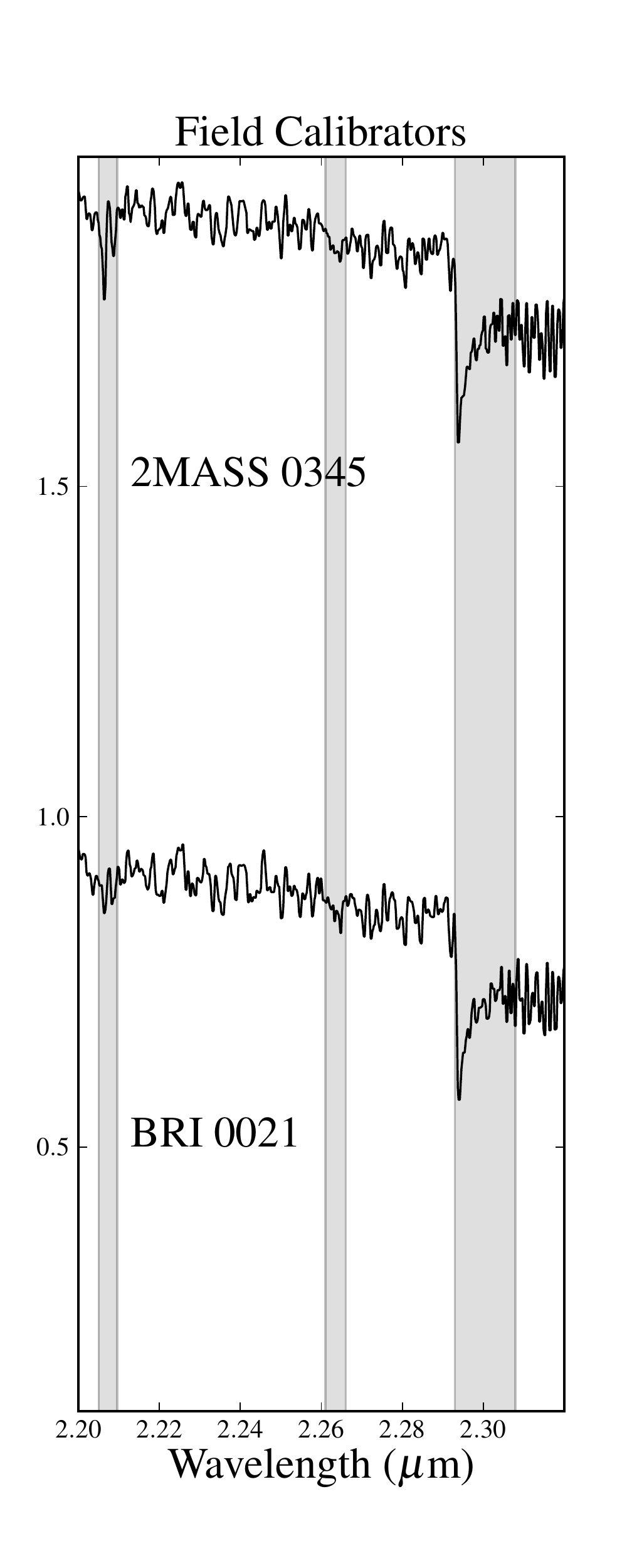}
   \end{subfigure} 
   \caption{Extracted spectra of the three types of calibrator. The normalised spectra are plotted with a constant integer offset. pEWs were calculated in the shaded regions. From left to right, these regions span the NaI doublet, the CaI triplet, and the first CO bandhead. The lower set of figures are expanded views of the shaded regions in the topmost figures.}
\end{figure*}
   
\section{Observations \& Data Reduction}
Observations of the ten calibrators were made over eight nights between 23 August 2008 - 29 October 2009. Observations of the PMOs were made on 16 October 2009. Observations were made using the 8m Gemini Telescope at Gemini North with NIFS~\citep{mcgregor03}. The latter was fitted with the K~band grating, centred at 2.2$\mu$m. Observations were normally made in an ABBA pattern to facilitate the removal of the sky background and dark current. Some characteristics of the brown dwarf calibrators and of the observed PMO, together with some observational notes including the S/N range of the NaI absorption feature for each object are shown in Table 1. These ranges were chosen as they are typical of the S/N for each observation.

\begin{table*}
\centering
\begin{minipage}{140mm}
\caption{Brown Dwarf Calibrator Sources and the PMO 152$-$717}  
\begin{tabular}{c c c c c c} 
\hline\hline
 & K~band Magnitude & Integration Time~(s) & S/N  & Aperture Size ($\prime\prime$)\\ [0.5ex] 
\hline
2MASS~0345$+$25 & 12.7 & 480 & 80$-$116 & 0.5 \\ 
BRI~0021$-$0214 & 10.6 & 480 & 158$-$184 & 0.6 \\
$\sigma$~Orionis~51 & 16.1 & 3600 & 29$-$59 & 0.23$^{1}$ \\
$\sigma$~Orionis~71 & 16.1 & 3600 & 34$-$73 & 0.26$^{1}$ \\
$\sigma$~Orionis~J053$-$024 & 16.2 & 3600 & 22$-$44 & 0.26$^{1}$ \\
KPNO-Tau 1 & 13.7 & 1000 & 96$-$123 & 0.5 \\ 
KPNO-Tau 4 & 13.3 & 1000 & 102$-$133 & 0.5 \\
KPNO-Tau 12 & 14.9 & 1800 & 52$-$119 & 0.5 \\
2MASS~0535$-$0546 & 13.8 & 500 & 84$-$131 & 0.45 \\
0457$+$3015 & 14.5 & 1800 & 63$-$129 & 0.44 \\ [1ex]

152$-$717 & 17.6 & 6300 & 4$-$8 & 0.2$^{1}$ \\ [1ex]
\hline
$^{1}$Observed using Adaptive Optics
\end{tabular}
\end{minipage}
\label{tab:1}
\end{table*}

Raw data were reduced using the GEMINI NIFS package within IRAF. The reduction was made in three steps: a baseline calibration to produce a shift reference file, a flat field file, a flat bad pixel mask file, a wavelength referenced arc file, and a spatially referenced ronchi file; a telluric calibration reduction to produce a 1D spectrum of the standard star to be used for telluric calibration of the science data; and a science data reduction to produce a 3D data cube which has been sky subtracted, flat fielded, cleaned of bad pixels, and telluric corrected.

The first two steps in the reduction process were completed by editing processing scripts supplied by the Gemini Observatory. The science reduction also largely followed a Gemini script. However, several additional steps were required to complete the reduction. In particular, the hydrogen Br$\gamma$ absorption line at 2.1661$\mu$m in the spectrum of the standard star chosen for the telluric calibration of each science object had to be removed, the modified spectrum then being divided by the star's blackbody spectrum and normalised before being divided into the extracted 1D science object spectrum to correct the latter for telluric absorption features.

The spectra were extracted using an aperture size of 1.5 times the full width at half maximum (FWHM) of each object, as determined from the dispersed images, see Table 1.

Observations in the K~band are susceptible to contamination by telluric OH sky lines. For the fainter objects observed here, the flux in the sky lines often varied sufficiently during the exposures that the lines were poorly subtracted in the reduction. To remove these lines as well as cosmic ray strikes and the general background, the data were processed using our own scripts to subtract the residual background along each column and interpolate across isolated pixels with highly anomalous counts. Care was taken to ensure that the scripts removed only noise features, using a comparison of the many image slices within each dispersed image to distinguish real features from noise. 

We used the IRAF task {\sc continuum} to normalise each science spectrum and remove the slope. We then fitted a second order chebyshev function to define a local pseudo-continuum. This pseudo-continuum set the flux level for measuring the equivalent widths of spectral features. While we experimented with higher-order fits, none differed significantly from the second-order fit. For this reason, the flux level of the pseudo-continuum was excluded as a source of error when we came to determine the uncertainties in our measurements. To obtain a representative pseudo-continuum, the fit excluded the NaI lines and the CO absorption bands starting at 2.29$\mu$m. Generally, the wavelength range for the fit was from  2.1$\mu$m $\rightarrow$ 2.2$\mu$m, and 2.21$\mu$m $\rightarrow$ 2.29$\mu$m. As a pseudo-continuum was used to measure these equivalent widths, they are more accurately referred to as pseudo-equivalent widths (pEWs). Spectral features were deblended using the IRAF task {\sc splot} and the pEWs derived by fitting Voigt functions to the deblended lines.

\section{Results}
\subsection{K~band Spectra}
The extracted 1D spectra of nine brown dwarf calibrators are shown in Figure 1. All spectra were smoothed using three pixel boxcars.

\begin{table*}
\centering
\begin{minipage}{140mm}
\caption{Brown Dwarf pseudo-Equivalent Widths (pEW)}  
\begin{tabular}{c c c c c}
\hline\hline
    & Age (Myr) & pEW NaI (\AA) & pEW CaI (\AA) & pEW CO (\AA) \\
\hline
2MASS~0345+25 & $\geq$1000 & 4.1$\pm$~0.2  & 1.1$\pm$~0.1  & 15.1$\pm$~0.2  \\
BRI~0021$-$0214 & $\geq$1000 & 2.2$\pm$~0.1  & 1.3$\pm$~0.1  & 18.8$\pm$~0.1  \\
$\sigma$~Orionis~51 & 3-7 & 1.0$\pm$~0.5  & 0.5$\pm$~0.3  & 9.3$\pm$~0.6  \\
$\sigma$~Orionis~71 & 3-7 & 1.8$\pm$~0.5  & $-$ & 8.4$\pm$~0.5  \\
$\sigma$~Orionis~J053$-$024 & 3-7 & 2.1$\pm$~0.6  & $-$ & 9.4$\pm$~0.5  \\
KPNO-Tau~1 & 1-2 & 1.8$\pm$~0.2 & 0.3$\pm$~0.1 & 12.4$\pm$~0.2 \\ 
KPNO-Tau~4 & 1-2 & 1.0$\pm$~0.1 & 0.6$\pm$~0.1 & 10.3$\pm$~0.1 \\
KPNO-Tau~12 & 1-2 & 1.2$\pm$~0.3 & $-$ & 9.8$\pm$~0.2 \\
2MASS~0535$-$0546 & 1-2 & 1.1$\pm$~0.1  & 0.9$\pm$~0.1 & 10.4$\pm$~0.1 \\
0457$+$3015 & 1-2 & 1.7 $\pm$~0.3 & 0.9 $\pm$~0.1  & 10.5 $\pm$~0.2 \\[1ex]
\hline
\end{tabular}
\end{minipage}
\label{tab:2}
\end{table*}

\begin{table*}
\centering 
\caption{Brown Dwarf Rotational Velocities}
\begin{tabular}{c c c}
\hline\hline
    & Rotational Velocity (v sin i) kms$^{-1}$ \\
\hline
2MASS~0345+25 & 25$^{1}$ \\
BRI~0021$-$0214 & 40$^{2}$ \\
$\sigma$~Orionis~51 & $-$ \\
$\sigma$~Orionis~71 & $-$ \\
$\sigma$~Orionis~J053$-$024 & $-$ \\
KPNO-Tau~1 & 5.5$^{3}$ \\ 
KPNO-Tau~4 & 10$^{3}$ \\
KPNO-Tau~12 & 5$^{3}$ \\
2MASS~0535$-$0546 & 10$^{4}$ \\
0457+3015 & $-$ \\[1ex]
\hline
$^{1}$~\citet{antonova08} \\
$^{2}$~\citet{reid99} \\
$^{3}$~\citet{mohanty05} \\
$^{4}$~\citet{reiners07}\\
\end{tabular}
\label{tab:3}
\end{table*}

\subsection{Equivalent Widths}
pEWs of the NaI doublet (2.206$\mu$m and 2.209$\mu$m) and of the CO band head starting at 2.294$\mu$m were obtained for all brown dwarf calibrators. pEWs of the CaI triplet (2.261$\mu$m, 2.263$\mu$m and 2.265$\mu$m) were obtained for both field dwarfs, one 3-7~Myr old object, and four 1-2~Myr old objects. These results are shown in Table 2. Table 3 shows, where known, the rotational velocities of the calibrators (obtained from the literature). Figure 2 shows the pEWs and their associated uncertainties. For comparison, Figure 2 also shows the NaI EWs of a sample of field dwarfs, M8 $\rightarrow$ M9.5, from the IRTF Spectral Library, contained in~\citet{cushing05} [C05] and~\citet{rayner09} [R09].

The pEW uncertainties in our sample of calibrator brown dwarfs were determined from the objects' dispersed 2D images. The image (before sky subtraction) was used to compute the total number of photons over the region containing the absorption feature. The width of this region was determined from the NIFS instrument profile and was found to be $\sim$2.2040$\mu$m - 2.2095$\mu$m for all our objects. The sky-subtracted 2D image of this region was used to determine the signal. The photon noise was then calculated using Poisson statistics.\footnote {We note that the VAR (variance) FITS extensions in the science data cannot be used for this purpose, owing to a problem with the GEMINI NIFS software package in {\sc IRAF}. The Gemini {\sc IRAF} team intends to fix this in a later release of the software.}

Our sample consists of objects in two known age groups, $\sim$1-2~Myr, $\sim$3-7~Myr, and those in the field. The age of the field objects is unknown, but they are significantly older than the objects in the known age groups. It would be interesting to examine the pEWs of gravity-sensitive features in objects with ages of a few 100~Myr. For example, the Pleiades cluster ($\sim$120~Myr) is known to contain a number of late~M/early~L dwarfs~(\citet{bihain10}, hereafter B10). (Note that later in this paper we examine an extended dataset, in which we derive the H$_{2}$(K) indices of several Pleiades objects.) We examined whether the CaI triplet and the CO trough at 2.29$\mu$m are also affected by surface gravity. For both these features, the pEWs of the two field dwarfs are larger than those of the other objects in our sample, particularly in the case of the CO trough, while the pEWs for these features among the two sets of younger calibrators are broadly similar.  

\begin{figure}
\centering
\begin{minipage}{\linewidth}
  \includegraphics[scale=0.3]{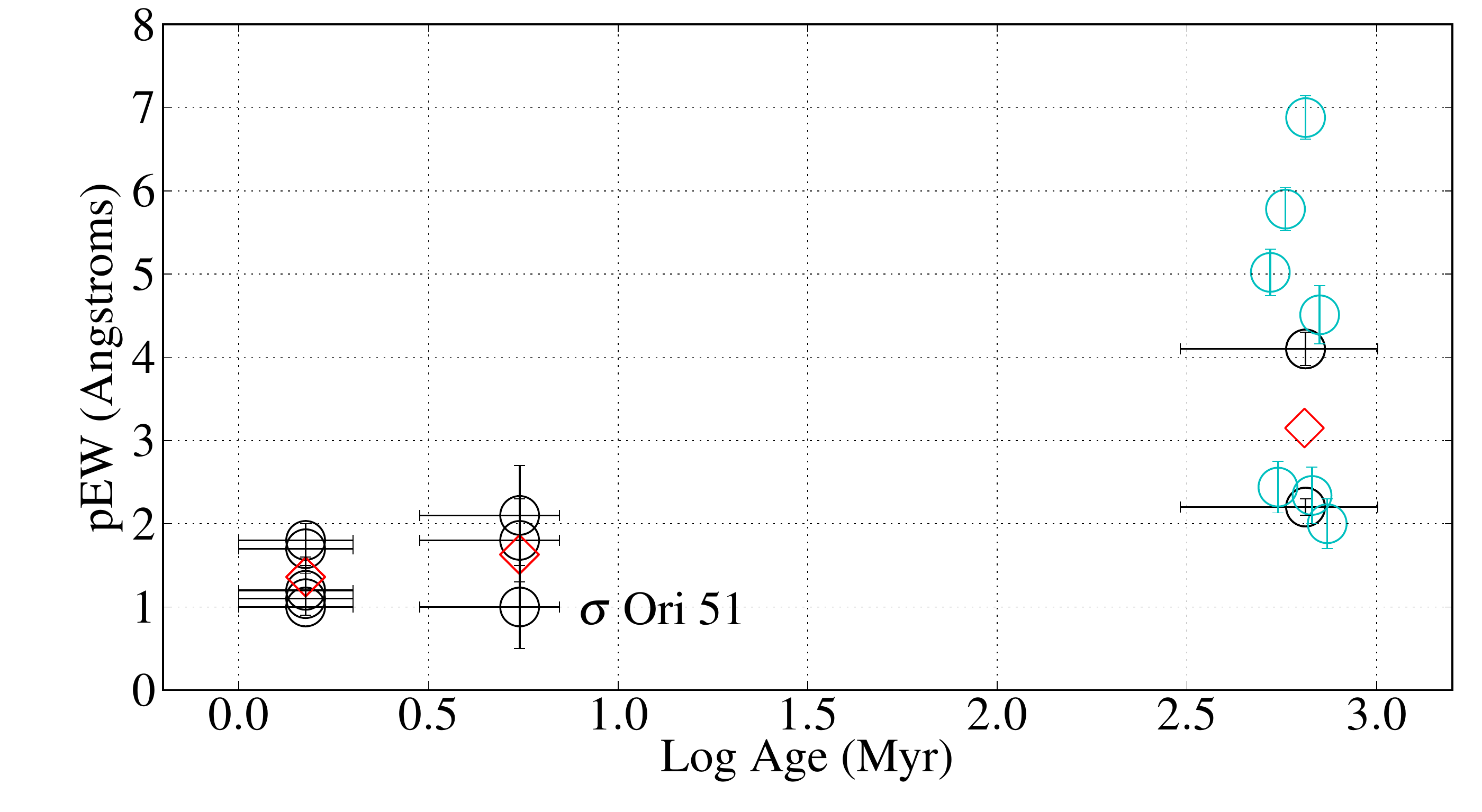} 
  \caption{NaI pEW as a function of age for our brown dwarf calibrators. The red diamonds are the mean pEWs of the brown dwarf calibrators at each age bin. The ages of the youngest objects are based on estimates in McGovern et al. 2004. The ages of the field dwarfs are speculative, but they are known to be much older than the other objects in this sample. $\sigma$ Orionis~51's pEW is more typical of a younger object. The cyan circles are the EWs of field dwarfs described in C05 and R09}.
  \label{fig:fig2}
\end{minipage}
\end{figure}

\subsubsection{Age Spreads}
Does the spread of values of pEW in each age category represent a real spread in ages? The spectrum either side of the NaI feature contains molecular absorption features which deepen with decreasing T$_{eff}$ (see Section 8.3). These features depress the local pseudo-continuum, leading to lower values of pEW. If our sample contained a wide range of spectral types, this is a possible cause of the spread of values of pEW. However, with one exception, our sample are spectral types M8.5$-$L0.  An ($I, I-J$) colour-magnitude study of 35 $\sigma$~Orionis cluster members found evidence of a spread in ages from 1-7~Myr~\citep{kenyon01}. A later survey of the $\sigma$~Orionis cluster found a large spread in measured values of Li pEW~\citep{kenyon05}, indicating a spread of ages. The authors did not reach any conclusions as to the cause of this spread in Li pEW, but noted that similar spreads in Li pEW have been observed among low mass objects in the Chamaeleon I cluster (\citealt{joergens01};~\citealt{natta04}). A study of the luminosity spread in the HRD of the ONC argued that there was little evidence to support age spreads greater than a few Myr~\citep{jeffries11}.

There is some debate whether star formation is a slow process, taking place over several Myr, or whether it takes only as much time as is required for a sound wave to cross a molecular cloud. The sound crossing time depends on the radius of the cloud and on the sound speed, but~\citet{elmegreen00} has suggested that star formation could take as little as 1~Myr or less. 

In the case of the $\sigma$ Orionis objects, $\sigma$ Orionis 51's pEW is 1.5$\sigma$ from the mean pEW of its siblings which might indicate an age spread. The scatter in the pEWs of the 1-2 Myr objects is also slightly greater than would be expected from the 1$\sigma$ uncertainties. However, the scatter in the data could have other causes. For example, magnetic activity can produce surface spots, so that in order to maintain its luminosity, an object has to increase in size. This results in the object having a lower surface gravity than it would have without any magnetic activity. An investigation of magnetic activity in field M dwarfs has suggested that activity peaks around M7, with significant activity continuing into later spectral types~\citep{hawley00}. Magnetic activity could be even greater among younger objects. A recent paper has suggested that the temperature reversal observed in 2MASS 0535$-$0546 could be due to magnetic fields inhibiting convection~\citep{mohanty12}.
 
While age spreads could explain the scatter in pEW values, our small sample size and the magnitude of the uncertainties on our data do not allow us to infer age spreads in young clusters. For example, the spread in data values could be due to observational scatter or may arise from differing formation mechanisms and accretion histories (B12). To settle the question of age spreads, we would need a sufficiently large sample of objects in order to undertake a rigorous statistical analysis.
 
\begin{table*}
\centering
\caption{Spectral Types}
\begin{tabular}{c c c c c}
\hline
 Object & Previous Spectral Type &\multicolumn{2}{c}{Spectral Type} & Revised Spectral Type\\
 & & WK & QK\\
\hline\hline
2MASS~0345$+$25$^{1}$ & L0 & $-$ & $-$ & $-$\\
BRI~0021$-$0214$^{1}$ & M9.5 & $-$ & $-$ & $-$\\
$\sigma$~Orionis~51 & M9 & $>$M9.5 & $>$M9.5 & L0$\pm$0.5\\
$\sigma$~Orionis~71 & L0 & M7.8 & M8.3 & M8$\pm$0.5\\
$\sigma$~Orionis~J053$-$024  & M9.5 & M8.1 & M9.3 & M8.75$\pm$0.5\\
KPNO-Tau 1 & M8.5 & M8.3 & M9.1 & M8.75$\pm$0.5\\ 
KPNO-Tau 4 & M9.5 & $>$M9.5 & $>$M9.5 & L0$\pm$0.5\\
KPNO-Tau 12 & M9 & M9.1 & M9.3 & M9.25$\pm$0.5\\
0457+3015 & M9.25 & M9.1 & M9.4 & M9.25$\pm$0.5\\ [1ex]
152$-$717 & M9 & $>$M9.5 & M9.3 & M9.5$\pm$1.0\\ [1ex]
\hline
$^{1}$Old field objects.
\end{tabular}
\label{tab:tab4}
\end{table*}

\subsection{Spectral Typing}
We took this opportunity to use our higher resolution spectra to re-examine the spectral typing of the younger objects in our sample, including the PMO. We used the WK and QK spectral indices previously defined by our group (L06), and the fits to these spectral indices used to characterise the GNIRS and NIRI data in L06 (W09). These indices provide spectral types consistent with the Luhman system of optical classification of spectra~\citep{luhman99}. QK is a reddening independent index, whereas WK requires knowledge of the extinction toward the source. For the PMO 152$-$717, the extinction A$_V$ = 3.7 (L06). Thus, we were  able to use the dereddened spectrum in determining its spectral type. For the $\sigma$~Orionis objects it is known that A$_V$ $\leq$ 1 mag, and, in most cases, changing A$_V$ by $\pm$ 2 mag alters the derived spectral type by $<$~0.5 types~\citep{bejar01}. Thus, we do not expect these spectra to be significantly affected by reddening. The same conclusion was made in the case of the Taurus association objects. While the WK index is sensitive to reddening, and therefore should be dereddened where possible, we have used it as a check on the spectral type derived from the more accurate QK index. We did not re-examine the spectral types of the field dwarfs in our dataset, as our spectral indices are only applicable to young objects. Our results are shown in Table 4. 

Spectral types were rounded to the nearest 0.25 subtype (which is the practice in the Luhman system). To assign a spectral type, we took the mean of the spectral types derived using the WK and QK indices. To determine the uncertainties in spectral types, we first measured the scatter in the fits to the WK and QK indices in the M8 to M9.5 interval in W09. This was added in quadrature to the mean average difference (0.3 sub-types) derived from the difference between the WK-based and QK-based spectral types, averaged over the 5 objects in this spectral type interval. Uncertainties were rounded up to the nearest 0.25 subtype. As the fits derived by W09 are uncalibrated beyond M9.5, when both WK and QK indices indicated a spectral type later than M9.5, we compared our spectra with the spectra of early L-type brown dwarfs in the Upper Sco association described in~\citet{lodieu08}, hereafter L08. As a result, we assigned a spectral type of L0 to these objects with an uncertainty of 0.5 spectral types.

We should qualify this re-examination of previously published spectra by saying that our classifications are based solely on our K~band spectra. Spectral classifications using other near-infrared or optical wavebands may produce slight differences for different objects. To examine this, we recalculated the spectral types of the Upper Sco objects in L08 using our indices and found that our indices tended to produce spectral types $\sim$0.5 spectral types earlier than those in L08. The variations are close to our estimated uncertainties, however, and we feel justified in this exercise based on the high quality of our data.

\begin{figure}
\centering
\begin{minipage}{\linewidth}
  \includegraphics[scale=0.3]{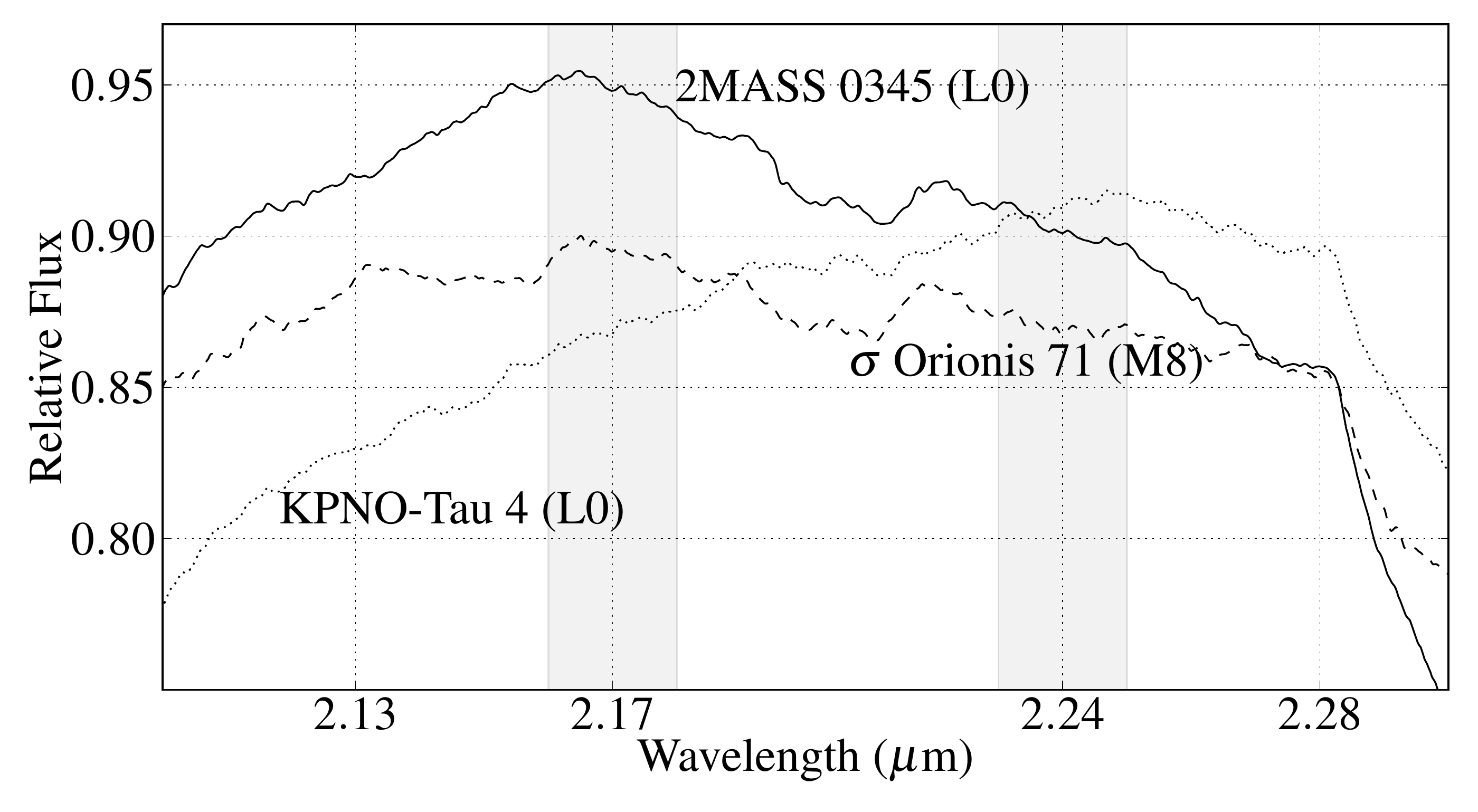} 
  \caption{Variations of the K~band slope with surface gravity are easily discernible in low resolution spectra. Here we plot smoothed versions of the NIFS spectra of 2MASS~0345 (field dwarf, solid line), $\sigma$~Orionis~71 (3-7~Myr, dashed line), and KPNO-Tau~4 (1-2~Myr, dotted line). The revised spectral types of these objects are indicated on the figure. The shaded areas are the wavelength regions over which the H$_{2}$(K) index is calculated.}
  \label{fig:fig3}
\end{minipage}
\end{figure}

\section{\textbf{H$_2$} Collision Induced Absorption in the K~band}
It has previously been observed that the F$_\lambda$ spectra of young, late M-type brown dwarfs have a flat maximum between 2.18$\mu$m and 2.28$\mu$m~\citep{luhman04a}, while mature field dwarfs of the same spectral type have a peak between 2.14$\mu$m and 2.18$\mu$m then decline between 2.18$\mu$m and 2.29$\mu$m (L06). In other words, the K~band spectra of younger objects appear redder than the K~band spectra of older objects. This change in the K~band spectrum is believed to be caused by increased collision-induced absorption (CIA) by H$_2$ in the atmosphere, owing to the increase in surface gravity with age (~\citealt{saumon12}, hereafter S12). We wanted to see whether this behaviour was evident in our sample, and, if it was, to quantify it.

\subsection{Measuring H$_2$ CIA}
We used the smoothed NIFS spectra of our brown dwarf calibrators to identify the locations of the peaks in their continua. Figure 3 demonstrates how smoothed spectra can be used to trace the shape of the K~band slope. 

Notice that the 1-2~Myr object has a positive slope in the region of interest, while the older objects have negative slopes. Note also that while $\sigma$~Orionis~71 is only a few Myr older than KPNO-Tau~4, its slope is already negative. We expect that as $\sigma$~Orionis~71 ages, its peak emission will not move far from its present wavelength value, but its K~band slope will become more negative, similar to the field dwarf calibrators. Thus, the slope of the K~band should be useful in differentiating very young pre-main sequence objects from older objects.

We found that the location of the peak in the continuum varies, depending on the surface gravity of the object. For objects with the largest surface gravity (the field dwarfs), the mean peak was at 2.17$\mu$m, while for objects with the smallest surface gravity (the 1-2~Myr objects), the mean peak was at 2.24$\mu$m.  Using the unsmoothed spectrum, the ratio of the median flux over a range of 0.02$\mu$m, centred at these wavelengths, defined an index, H$_{2}$(K), 

\begin{equation}
H_{2}(K)=\frac{F_\lambda(2.17\mu\rm{m})}{F_\lambda(2.24\mu\rm{m})} 
\end{equation}

For young objects, with a positive, or at least flat slope, between these limits, the index returns a value $\le$ 1. For older objects, whose slope is negative, the index is $>$ 1. 

The H$_{2}$(K) indices for our sample of brown dwarf calibrators are shown in Figure 4 and in Table 5. They are consistent with the results for the correlation of age with pEW shown in Figure 2. The uncertainties on the indices were determined using the standard error on the mean in the 0.02$\mu$m intervals over which the indices were calculated.

\begin{figure}
\centering
\begin{minipage}{\linewidth}
  \includegraphics[scale=0.3]{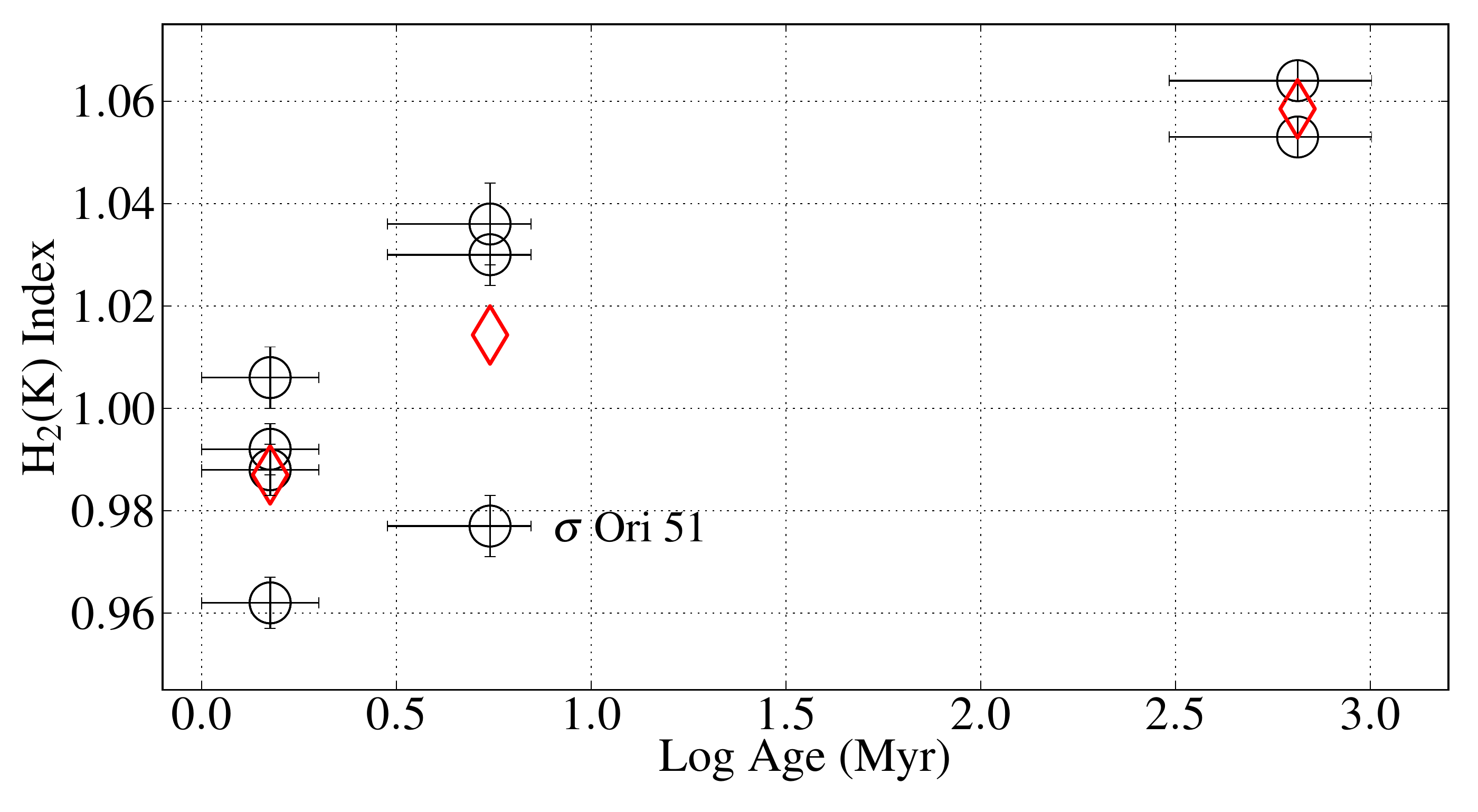} 
  \caption{H$_{2}$(K) index as a function of age for our brown dwarf calibrators. The red diamonds are the mean H$_{2}$(K) indices for each age bin. The ages of the objects are as described in Figure 2. Again, notice the position of $\sigma$~Orionis~51.}
  \label{fig:fig4}
\end{minipage}
\end{figure}

\begin{table*}
  \centering
  \caption{H$_{2}$(K) Indices, Means and Standard Deviations}\label{tab:5}
  \smallskip
  \begin{minipage}{85mm}
  \begin{tabular}{c c c c c c}
    \hline\hline
    & H$_{2}$(K) Index & ${\lambda_{F_{MAX}}}$ ($\mu$m) & $\bar{x}_{\rm{H_{2}(K)}}$ & $\sigma$\\
    \hline
    2MASS~0345+25 & 1.053$\pm$0.004  & 2.16 & 1.059 & 0.008\\
    BRI 0021$-$0214 & 1.064$\pm$0.004 & 2.16 & & \\
    \hline
    $\sigma$~Orionis~51 & 0.977$\pm$0.006 & 2.26 \footnote{Note that the peak flux occurs at a wavelength
    more typical of a younger object.} & 1.014 & 0.032\\
    $\sigma$~Orionis~71 & 1.030$\pm$0.006 & 2.16 & &\\
    $\sigma$~Orionis J053$-$024 & 1.036$\pm$0.008 & 2.19 & &\\
    \hline
    KPNO-Tau~1 & 0.992$\pm$0.005 & 2.23 & 0.987 & 0.018 \\ 
    KPNO-Tau~4 & 0.962$\pm$0.005 & 2.25 & & \\
    KPNO-Tau~12 & 1.006$\pm$0.006 & 2.17 & &\\
    0457+3015 & 0.988$\pm$0.005 & 2.25 & & \\
    \hline
   \end{tabular}\par
   \vspace{-0.75\skip\footins} 
   \renewcommand{\footnoterule}{}   
  \end{minipage}
\end{table*}

Among the calibrators, one 1-2~Myr object (KPNO-Tau~12) has a H$_{2}$(K) value $>$ 1, implying a negative slope (if only slightly). Also, $\sigma$~Orionis~51 has an unusually low H$_{2}$(K) value. This is consistent with this object's NaI pEW. One should not use a single age/gravity indicator to infer the values of either of these quantities in one object. However, $\sigma$~Orionis~51 has two indicators of low gravity. While this spread in data values may be due to differing formation mechanisms and/or accretion histories, it can also result from an age spread in the $\sigma$~Orionis cluster.

Although the ONC PMO 152$-$717 is not one of our brown dwarf calibrators, the object's data were obtained in this observation and we think it is appropriate to discuss the results here. While the S/N of the extracted spectrum was too low to allow us to measure the NaI pEW, we were able to calculate its H$_{2}$(K) index. With a value of 0.914$\pm$0.013, 152$-$717's H$_{2}$(K) index is consistent with 152$-$717 being a low gravity object. Its value is also consistent with the values of the H$_{2}$(K) indices for the other ONC PMOs in the extended dataset (see Section 7 and Table 6). This highlights an important benefit of this method of constraining the surface gravity, and hence age, of these objects. Since the H$_{2}$(K) index is obtained by sampling the slope of the K~band, this technique allows us to obtain indications of surface gravity using spectra which are too noisy or where $\lambda\//\Delta\lambda$ is too low to resolve narrow spectral features.

\subsection{The Model H$_{2}$(K) Indices}

\begin{figure*}
\centering
\includegraphics[scale=0.45]{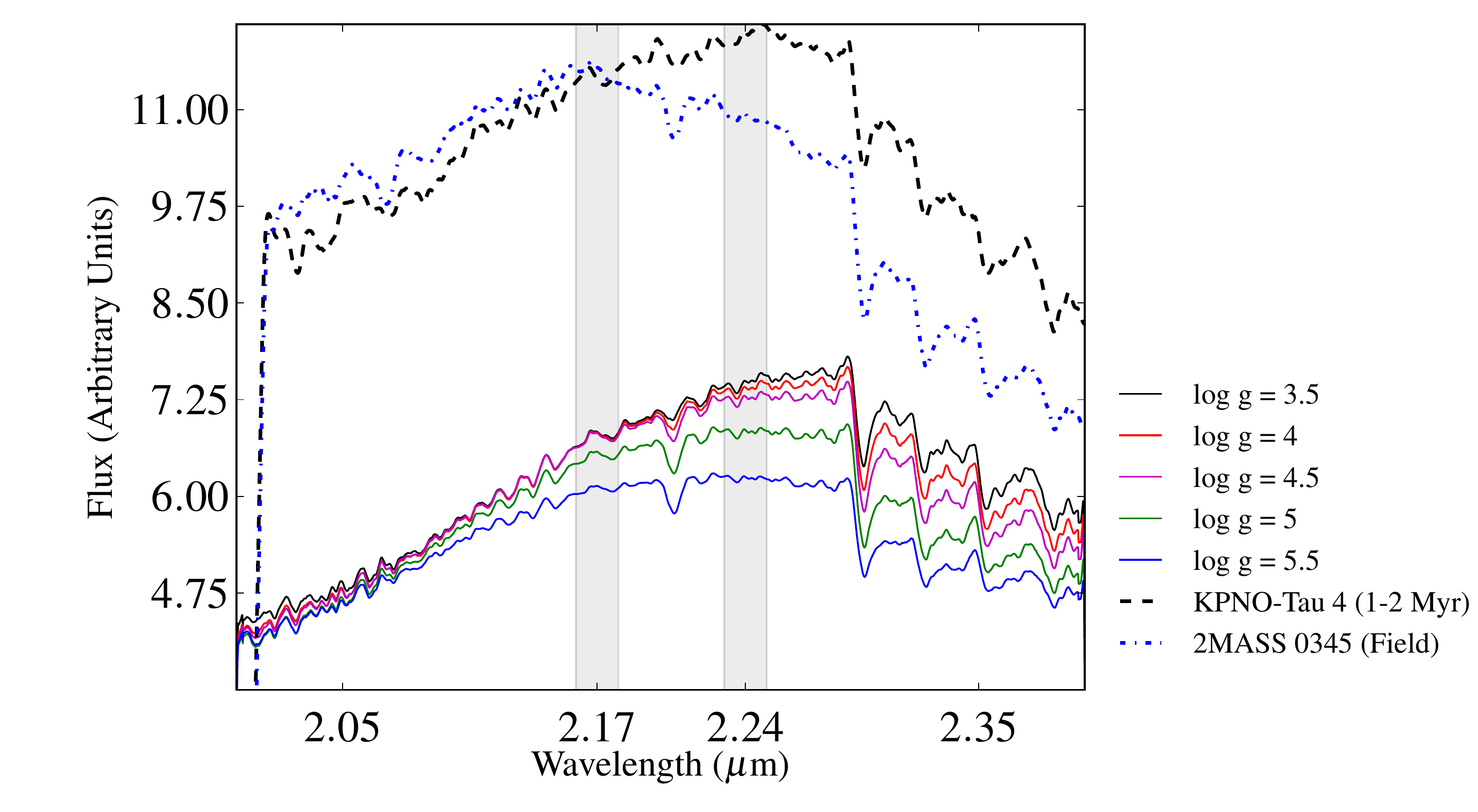} 
\caption{Theoretical models made using new calculations of H$_2$ CIA show the K~band slope decreasing	with increasing surface gravity. The temperature for these models was set at 2200K (the typical T$_{eff}$ of an M9.5$-$L0 dwarf). Overplotted are two spectra from our dataset of brown dwarf calibrators, a young Taurus object, and an older field dwarf. The shaded regions are as those described in the caption to Figure 3. All spectra have been smoothed to aid separation. (Models supplied by Didier Saumon and Mark Marley).}	
\label{fig:fig5}
\end{figure*}

\subsubsection{Modelling H$_2$ CIA in the K~band}
The objects in our sample of brown dwarf calibrators are of a similar spectral type. They differ most significantly in age, and therefore in surface gravity. Molecular hydrogen is the most abundant constituent of brown dwarf atmospheres and H$_2$ CIA is positively correlated to density. At any given metallicity, objects with greater surface gravity will show enhanced H$_2$ CIA. 

H$_2$ absorption in the K~band has been examined before (\citealt{tokunaga99}, hereafter T99). T99 defined two parameters. The first parameter, K1, measures the slope of the spectrum between 2.00$\mu$m and 2.14$\mu$m.  The second parameter, K2, measures the strength of H$_2$ absorption between 2.14$\mu$m and 2.24$\mu$m. T99 suggested that the K2 parameter could be used to estimate an object's T$_{eff}$. They did not apply this index to constrain an object's age other than to say that the K2 parameter was generally negative for field dwarfs. In a subsequent survey of young cluster members, the K2 parameter was typically positive (L08). 

To further examine the effect of H$_2$ CIA on the shape of the K~band, we used newly computed models of H$_2$ CIA in the atmospheres of late-type brown dwarfs to generate synthetic spectra at T$_{eff}=$~2200K (the typical T$_{eff}$ of an M9.5$-$L0 dwarf), over a range of surface gravities from  log~g~$=$~3.5$\;\rightarrow\;$5.5 [S12]. The results are shown in Figure 5. These Saumon \& Marley models assume that the dwarfs' atmospheres are cloudy $(F_{sed}=2)$ and of solar metallicity.

The models show a change of slope in the region 2.16$\mu$m - 2.28$\mu$m, with the slope decreasing with increasing gravity. They predict that the rate of change of the slope increases with increasing gravity. They also predict that the spectra should show greatest sensitivity to the increasing contribution of CIA H$_2$ to the opacity of the atmosphere in the K~band at log g~$>$~4.5, rather than at the lower gravities expected in pre-main sequence clusters. The data imply a stronger dependence of CIA H$_2$ opacity on gravity than the models predict, particularly at log g~$<$~4.5. 

In Figure 6, we used Lyon DUSTY isochrones~(\citet{chabrier00};~\citet{baraffe02}) to estimate the surface gravities of our brown dwarf calibrators (see Section 8.1). We took the mean surface gravity for each age group of brown dwarf calibrators, and assigned to that gravity the mean H$_2$(K) index for that age group. We then plotted the Saumon \& Marley 2200K model H$_{2}$(K) indices as a function of surface gravity, fitted a simple polynomial to the data points and interpolated between the points to determine the values of the model H$_{2}$(K) indices at the surface gravities of the brown dwarf calibrators. We repeated the procedure using an extended dataset (seen Section 7), and found a similar relationship between data and models. 

The disparity between models and data at lower gravities is clear. The changes in the H$_2$(K) index at lower gravities are much larger for the data than for the models. It may be that the models overestimate the significance of dust opacity at these lower gravities. It is possible that young objects can have higher T$_{eff}$ than field dwarfs of the same spectral type~\citep{luhman03b}. This higher T$_{eff}$ may prevent dust condensation, allowing H$_2$ CIA opacity to play a greater role. 

The degeneracy in the data at log g $\sim$5 may be due to the very low resolution (R~$\approx$~50) of the data point at log g~$=$~4.90 (Teide 1, a Pleiades object from the extended dataset (see Section 7)). It is also possible that the H$_2$(K) index becomes saturated at log g~$\geq$~5. While this behaviour is not shown by the model spectra, we note that the other Pleiades objects in the extended dataset, HII 1248B and BPL 62, have H$_2$(K) indices of 1.068 and 1.138 respectively. We assume that these objects have similar surface gravities. If the H$_{2}$(K) index does saturate at these values of log g, this would imply an upper limit on the age sensitivity of the index.

Figure 7 shows a strong correlation between the NaI pEWs and the H$_2$(K) indices calculated from our dataset of brown dwarf calibrators and from the Saumon \& Marley 2200K models. The strength of gravity-sensitive neutral alkali metal lines is a well-proven method in establishing the ages of late-type M~dwarfs. This correlation suggests that the H$_2$(K) index is at least as good an age indicator. (See Section 9 for a more detailed comparison of the two age indicators.) 

\begin{figure}
\centering
\includegraphics[scale=0.3]{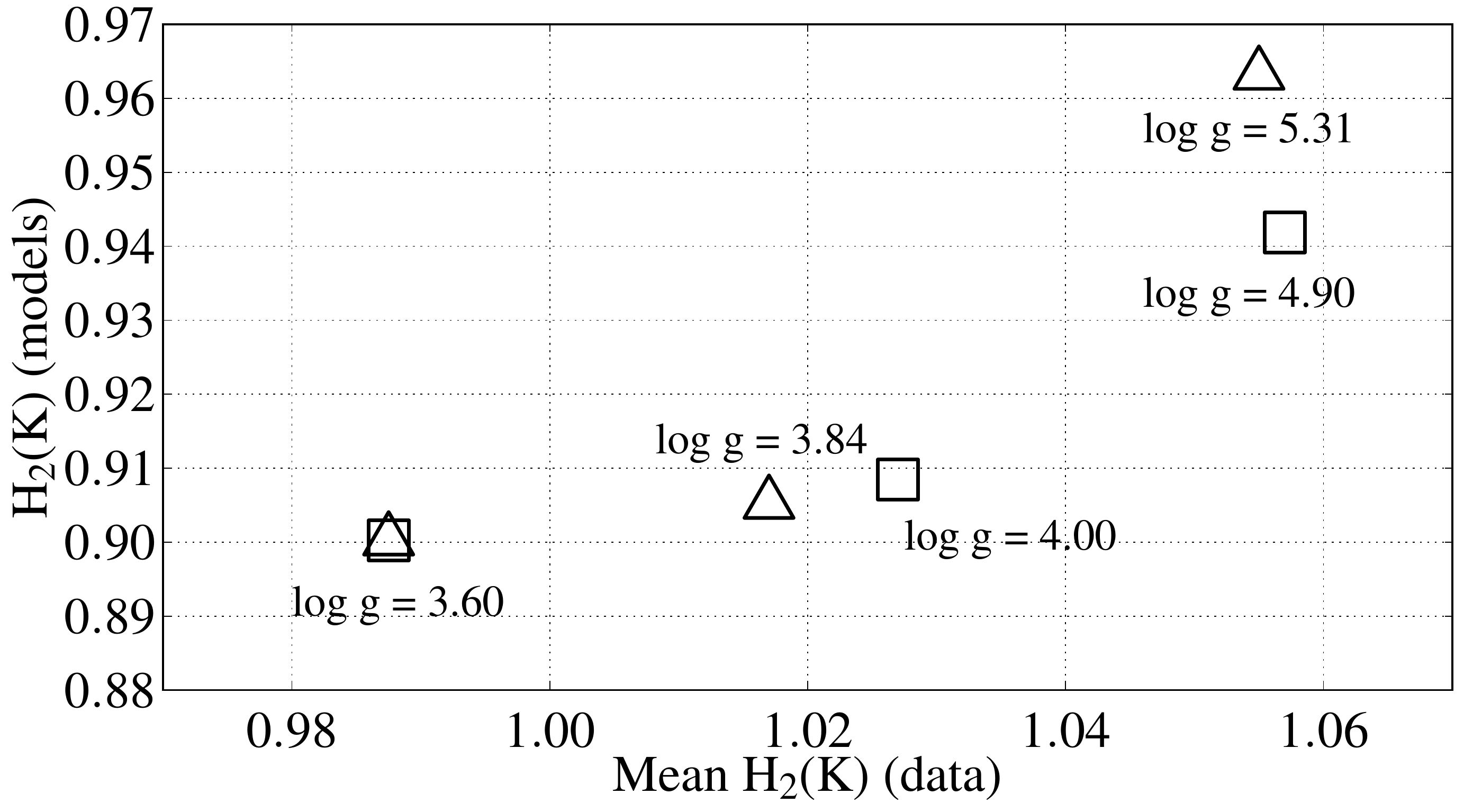} 
\caption{The relationship between the mean H$_{2}$(K) indices of our original dataset (triangles), the extended dataset (squares) and the model H$_2$(K) indices. The observed dependence of the H$_{2}$(K) index as a function of gravity appears to be substantially steeper than predicted by the models at low surface gravities.}	
\label{fig:fig6}
\end{figure}

\begin{figure}
\centering
\includegraphics[scale=0.30]{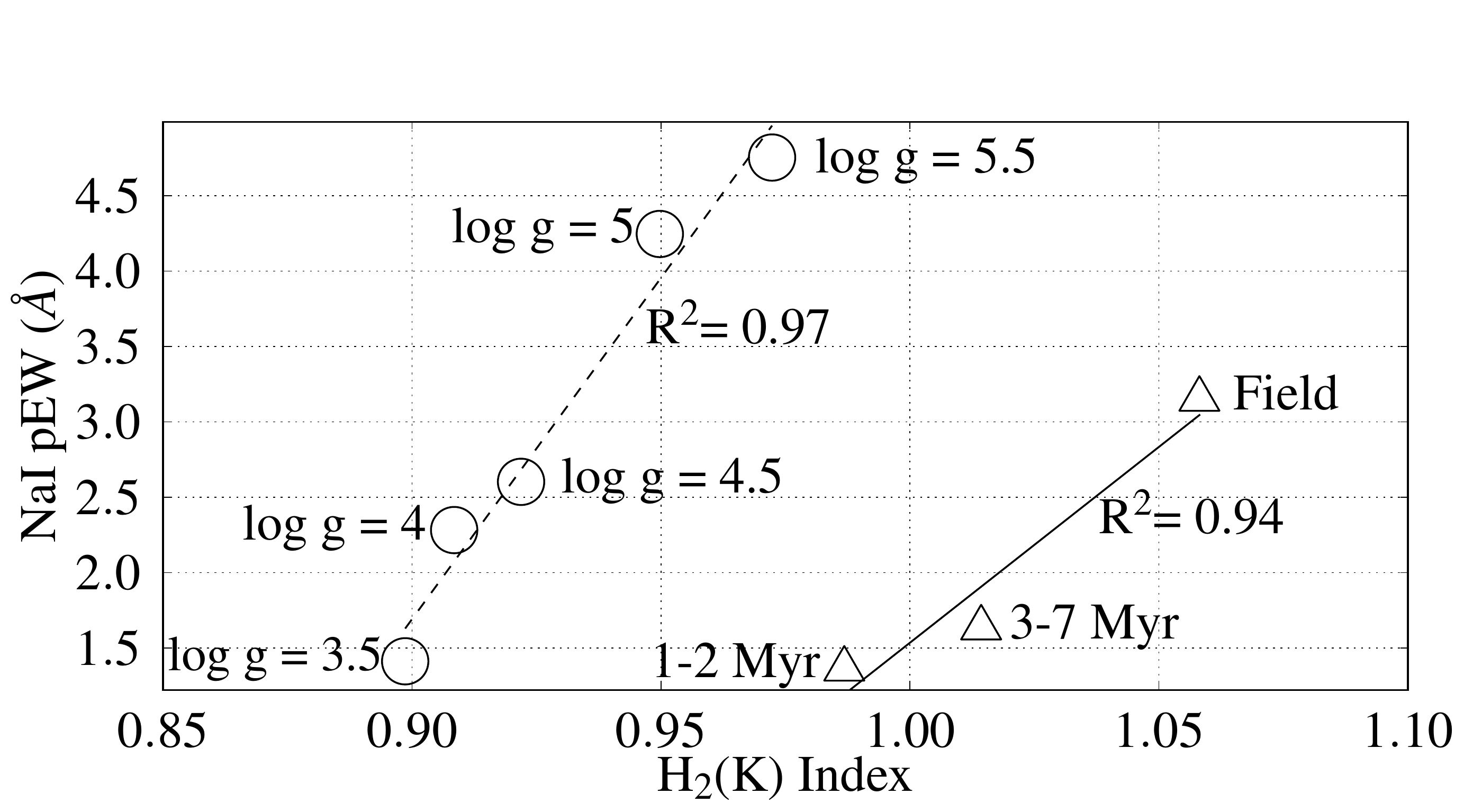} 
\caption{The correlation between NaI pEWs and H$_{2}$(K) indices for our original dataset (triangles, solid line) and the Saumon \& Marley model spectra (circles, dashed line). Note that we have used the means of these values for the original dataset.}	
\label{fig:fig7}
\end{figure}

\subsection{Testing the H$_{2}$(K) Index}
\subsubsection{The Effect of Dust and Extinction}
In cases of unresolved binarity, emission from a disc, non-sphericity, etc, the K~band may exhibit a flux excess, affecting the H$_{2}$(K) index. A common cause of K~band flux excess in young objects is emission from a dusty accretion disc. To examine the effect of dust emission on the K~band of a typical 1-2~Myr calibrator (KPNO-Tau~4), it was assumed that half the calibrator's flux was produced by dust, simulated by a blackbody function. This is the level of contamination we would expect to find in the K band spectra of very dusty objects. We performed two simulations of the blackbody function, at 900K and 1200K respectively, see Figure 8. 
The 1200K blackbody spectrum peaks at $\sim$2.4$\mu$m. The combined 1200K blackbody and science object spectrum (dotted line) peaks at $\sim$2.25$\mu$m, relatively close to KPNO-Tau~4's peak flux (solid line). The blackbody spectrum is flatter near its peak so adding a large component of this flux to the spectrum of KPNO-Tau~4 causes the latter to similarly flatten, increasing its H$_{2}$(K) index. The 900K blackbody spectrum peaks at $\sim$3.2$\mu$m. The combined 900K blackbody and science object spectrum (dashed line) peaks at $\sim$2.28$\mu$m. This has the opposite effect, steepening KPNO-Tau~4's spectrum and decreasing its H$_{2}$(K) index.

The H$_{2}$(K) index samples the K band flux, the region of the near infrared most susceptible to contamination by dust emission. While the effect of contamination by dust emission is somewhat mitigated by the narrow baseline (0.07$\mu$m) over which the index is calculated, the emission from the dusty objects simulated in Figure 8 does alter the H$_{2}$(K) index of KPNO-Tau~4 by $\sim\pm$0.02. While this is significant, we note that the presence of such large amounts of emission from a dusty disc becomes obvious in the spectrum (see Figure 8), so that such objects can be excluded when calculating an average H$_{2}$(K) index for a population.

The H$_{2}$(K) index will be affected by extinction. For example, assuming $A_{\textit{v}}=~5$ magnitudes, the H$_{2}$(K) index of a noiseless model spectrum decreases from 0.97 to 0.95. However, in most cases it will be possible to deredden a spectrum accurately, thereby minimising the effect.
 
\begin{figure}
\centering
\includegraphics[scale=0.3]{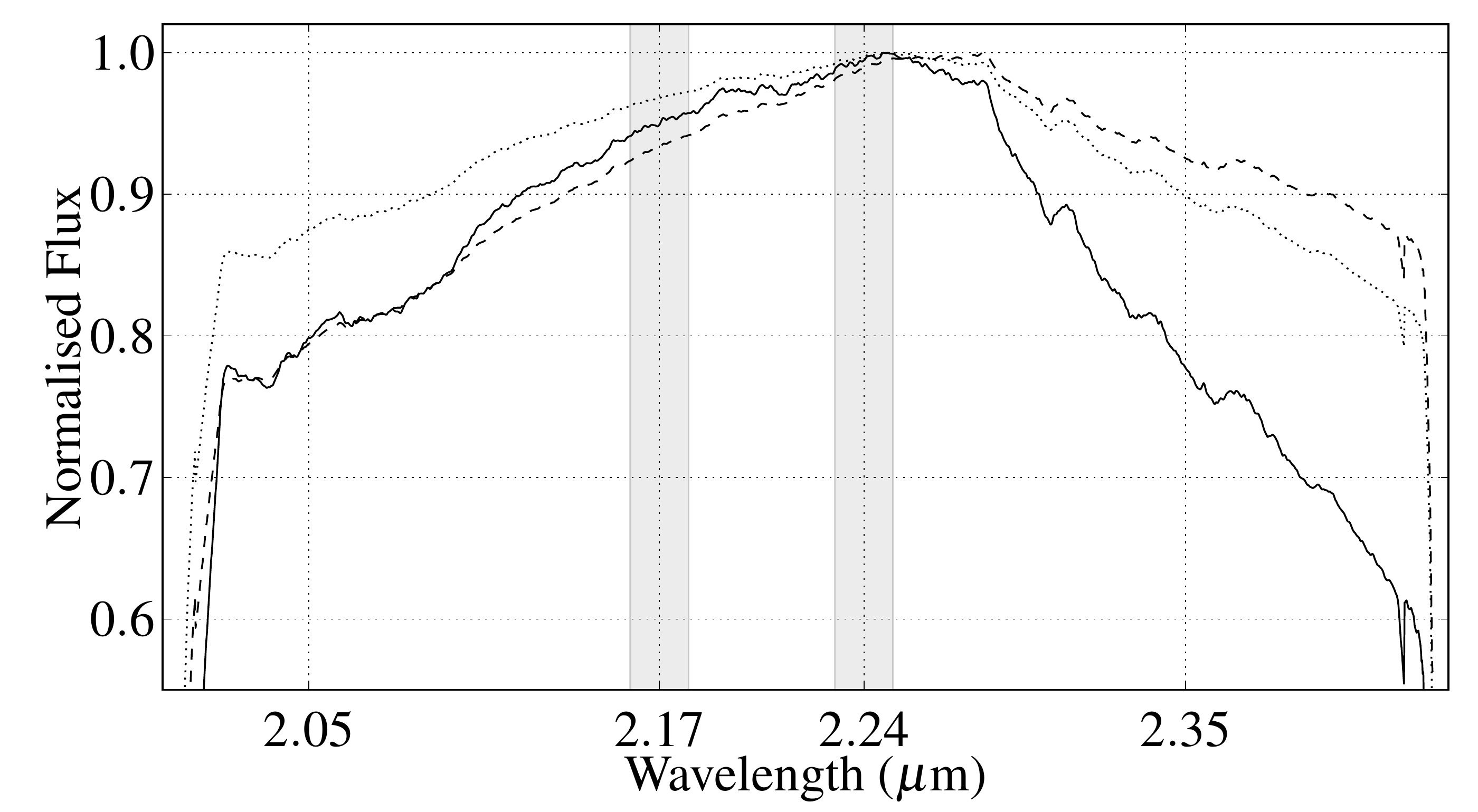}  
\caption[Dust Veiling]{\footnotesize The effect of dust emission on K band spectra. Emission from blackbodies at 900K and at 1200K are added to the smoothed spectrum of KPNO-Tau 4 (dashed line and dotted line, respectively) and contrasted with the original spectrum (solid line). The shaded areas are as described in Figures 3 and 5.}
\label{fig:fig8}
\end{figure}

\begin{figure}
\centering
\includegraphics[scale=0.3] {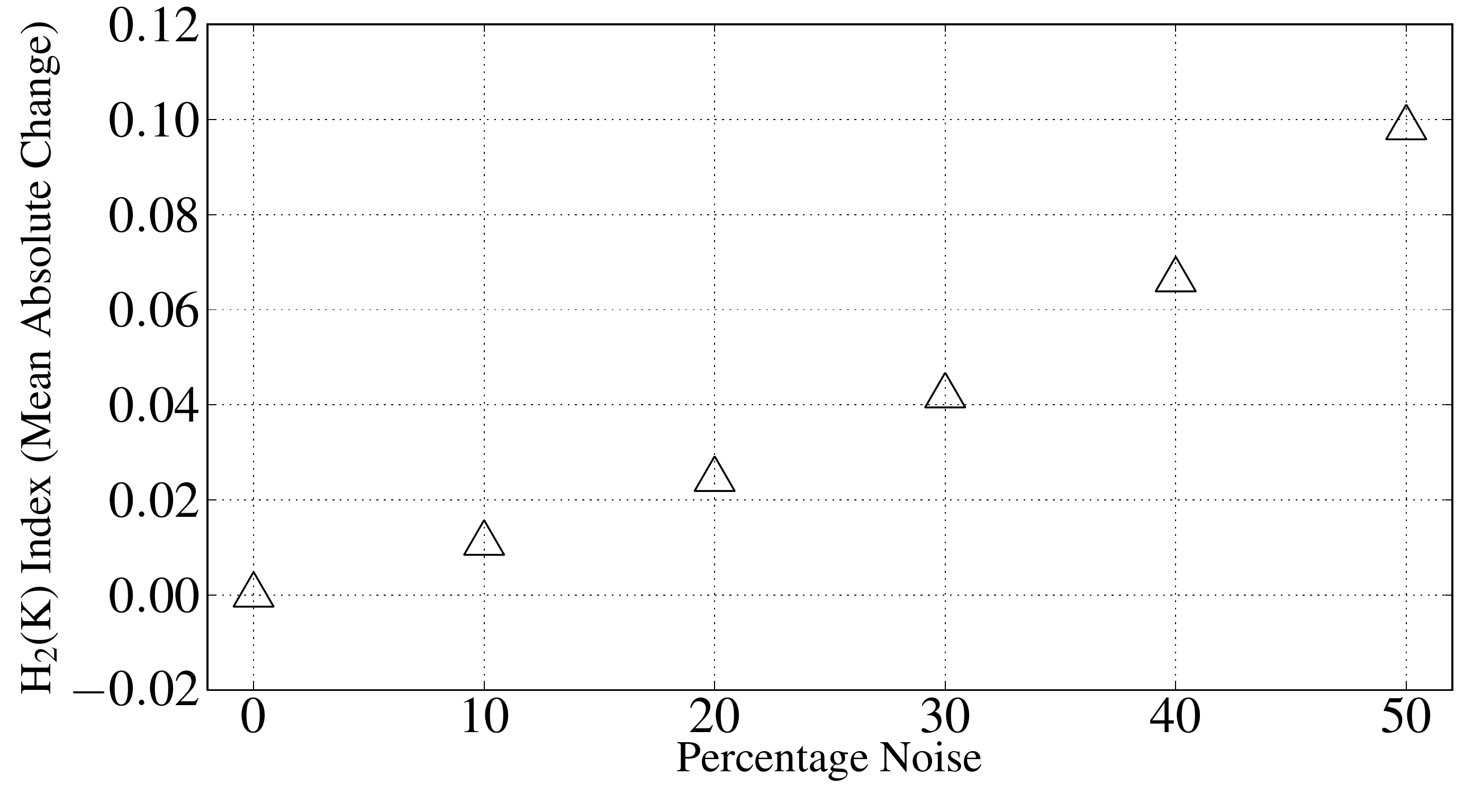} 
\caption{The effect of noise on the H$_{2}$(K) index. Randomly generated noise was added to a model spectrum ($T_{eff}$ 2200K, log g = 5.5).}
\label{fig:fig9}
\end{figure}

\subsubsection{The Effect of Noise}
We carried out a Monte Carlo analysis to examine the effect of noise on the H$_{2}$(K) index. We chose a noiseless model spectrum and added successively larger fractions of noise. The noise was added randomly to each pixel of the model spectrum, at the spectral resolution of our NIFS spectra. After 10${^4}$ iterations of each stage we measured the mean absolute change in the H$_{2}$(K) index. The results are shown in Figure 9. The change in the H$_{2}$(K) index exceeds $\pm$~0.02 when noise comprises $>$ 20\% of the flux.

\subsubsection{The Effect of Metallicity}
Our discussion on 2MASS~0535$-$0546 (see Section 8.1.3) concludes that alkali metal absorption features in mid to late M dwarfs are almost pure measures of surface gravity, with very little dependence on temperature. While surface gravity is probably the largest factor in determining the shape of the K~band slope, metallicity is likely to have some effect. Reducing the metallicity removes H$_2$O and CO opacity from the K~band while other transmission windows (YJH) become more transparent since they don't have as much H$_{2}$ CIA absorption. Thus, the K~band flux decreases relative to YJH (D.Saumon, private communication). Increasing the metallicity has the opposite effect.

In summary, K~band flux is reduced in the case of low metallicity or high gravity, and enhanced where there is low gravity or high metallicity~\citep{leggett07}.

While variations in metallicity may affect the K~band slope, the objects we have examined in this paper reside in regions of solar metallicity where variations in metallicity are negligible~\citep{santos08}.

\subsubsection{The H$_{2}$(K) index as a Function of Spectral Type}
We looked at variations in the H$_{2}$(K) index due to the spectral type of the objects in our complete dataset. In Figure 10 we plot these objects' H$_{2}$(K) indices as a function of their spectral types. The plot shows no clear trend and a simple linear fit to the data has a correlation coefficient of $<$0.1.

\begin{figure}
\centering
\includegraphics[scale=0.30]{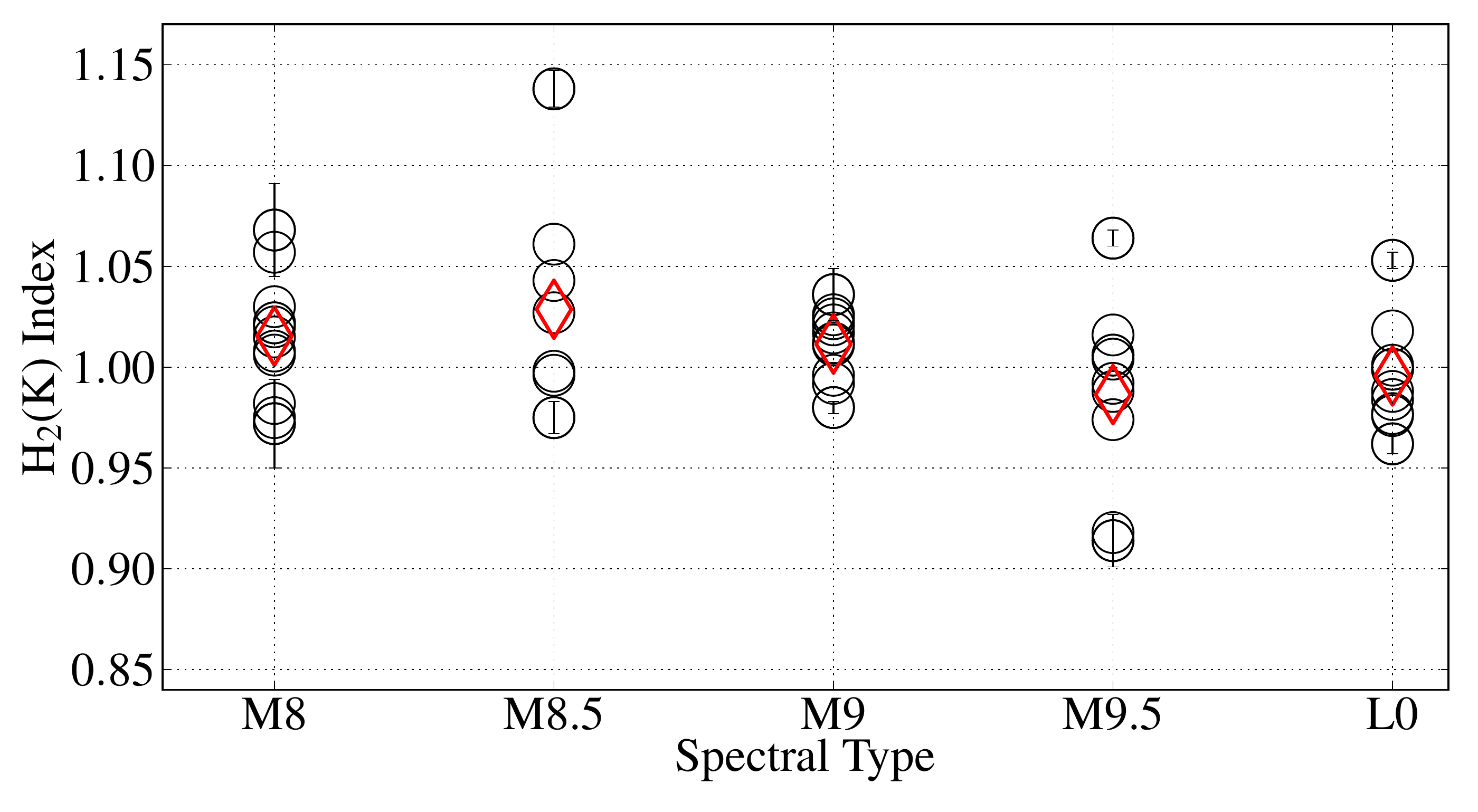} 
\caption{The H$_{2}$(K) indices of the complete dataset, plotted as a function of spectral type. Red diamonds are the mean H$_{2}$(K) index values. Errorbars are shown for the outlying objects at each spectral type.}
\label{fig:fig10}
\end{figure}

\section{Extending the Dataset}
To further test the correlation of the H$_{2}$(K) index with surface gravity, and hence age, we examined an extended dataset from the literature, containing field dwarfs and objects in clusters with constrained ages. Although we have not found the H$_{2}$(K) index to be sensitive to spectral type over the range M8 to L0, in our dataset of calibrator brown dwarfs, objects in the extended dataset were selected to cover the same range of spectral types to keep the analyses comparable. We should also point out that despite constraints, the ages of these clusters are not certain and our analysis should be interpreted in the light of this fact. For example,~\citet{pecaut12} have argued that Upper Sco is 11 Myr old, compared to previous estimates of $\sim$5 Myr. Pecaut et al's argument is compelling and we have used their revised age for Upper Sco in our extended dataset.

We were particularly keen to obtain spectra from clusters with ages $\geq$ 100 Myr since the older objects among our brown dwarf calibrators are field dwarfs of indeterminate age. We searched the literature for good quality K band spectra of members of clusters such as the Pleiades, Hyades and Praesepe, but found spectra of only three objects in the Pleiades and none at all in the Hyades or Praesepe. Our attempt to constrain the H$_{2}$(K) index has been hindered by this lack of high quality K band spectra of objects in older clusters. 

The extended dataset comprises 

\begin{list}{$\circ$}{} 
\item an M8.5 field dwarf described in~\citet{geballe02} [G02];
\item an M8.5 field dwarf described in~\citet{leggett01} [L01];
\item seven field dwarfs, M8 $\rightarrow$ M9.5, obtained from the IRTF Spectral Library, described in C05 and R09;
\item Teide 1, an M8 dwarf in the $\sim$120 Myr Pleiades cluster, described in B10;
\item HII 1248 B, an M8 $\pm$ 1 substellar companion to the Pleiad H11 1348 A, described in~\citet{geissler12} [G12];
\item BPL 62, an M8.3 dwarf in the Pleiades cluster. This is an unpublished spectrum taken by D. J. Pinfield. The object is described in~\citet{pinfield00} [P00] and~\citet{pinfield03} [P03]; 
\item GSC08047 B, an M9.5 dwarf in the $\sim$30 Myr Tuc-Hor association, described in~\citet{patience12} [P12]; 
\item 13 M8 $\rightarrow$ L0 brown dwarfs in the $\sim$11 Myr old Upper Sco association, described in L08;
\item TWA~5B, an M8/M8.5 dwarf in the $\sim$8 Myr TW Hya association (TWA), described in~\citet{neuhauser09} [N09];
\item DENIS J124514.1-442907 and 2MASSW J1139511-31592 (2MASS 1139-3159), M8 $\rightarrow$ M9.5 objects in TWA, described in~\citet{looper07} [L07];
\item 2MASS J12073346-3932539 (2M1207A), a TWA object, described in P12; 
\item CT Cha B, an M8 $\sim$2 Myr object described in~\citet{schmidt08} [S08];
\item four brown dwarfs in the Taurus association, M8.25 $\rightarrow$ M9.25, described in~\citet{luhman04b} [L04b];
\item LRL 405, an M8 $\sim$1-2 Myr object in the IC 348 nebula, described in~\citet{muench07} [M07];
\item four ONC PMOs, M8 $\rightarrow$ M9.5, described in W09;
\end{list}
The results are shown in Figure 11 and Table 6.

\begin{figure}
\centering
\includegraphics[scale=0.3]{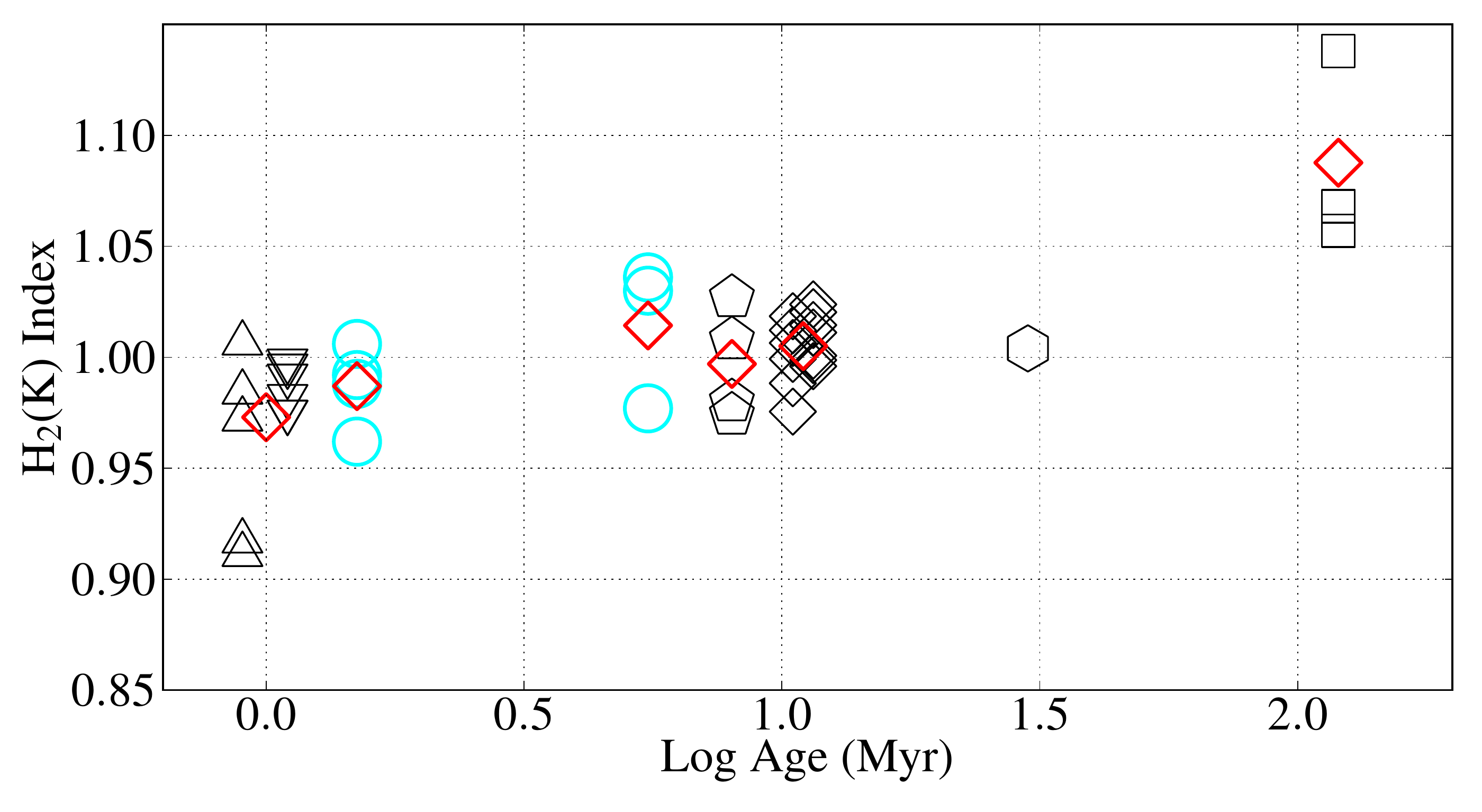}
\caption{The H$_{2}$(K) indices of the extended dataset, plotted as a function of age. Triangles are the GNIRS PMOs and 152$-$717, inverted triangles are the Taurus objects from L04b, diamonds are the Upper Sco objects described in L08, the pentagons are the TW Hya objects, the hexagon is GSC08047 B and the squares are the Pleiades substellar objects. The field dwarfs' H$_{2}$(K) indices have not been plotted as the ages of these objects are uncertain.  Red diamonds are the mean values at each age bin. Note that the rate of increase in the index is greatest for the youngest objects, as would be expected for rapidly contracting objects. The Taurus and ONC objects probably have similar ages so are shown offset slightly either side of 1~Myr, and share a common mean. For clarity, the Upper Sco objects are offset 0.5~Myr either side of 11~Myr. The cyan circles are the H$_{2}$(K) indices of our brown dwarf calibrators.}
\label{fig:fig11}
\end{figure}

\begin{table*}
\scriptsize
\begin{minipage}{140mm}
\caption[Extended Dataset $\rm{H_{2}(K)}$ Indices]{\footnotesize Extended Dataset $\rm{H_{2}(K)}$ Indices, Peak Flux Wavelength, Means, and Standard Deviations. (All USco and GNIRS objects have been dereddened.)}
\begin{tabular}{l l l l l l l l l l l}
\hline\hline
\centering
& Ref & Sp Type & $\sim\;$Age (Myr) & H$_2$(K) Index & $\rm{\lambda_{max}}$ ($\rm{\mu}$m) & $\bar{x}_{\rm{H_{2}(K)}}$ & $\sigma$ \\ 
\hline
SDSS~2255 & G02 & M8.5 & Field & 1.061$\pm$0.012 & 2.17 & 1.030 & 0.016 & &\\
T~513 & L01 & M8.5 & Field & 1.043$\pm$0.003 & 2.15 & & & &\\
2MASS J07464256+2000321AB & C05 & L0.5 & Field & 1.046$\pm$0.004 & 2.17 & & & &\\
DENIS-P J104814.7-395606.1 & R09 & M9 & Field & 1.026$\pm$0.004 & 2.17 & & & &\\
LP 944-20 & C05,~R09 & M9 & Field & 1.017$\pm$0.003 & 2.15 & & & &\\
LHS 2065 & R09 & M9 & Field & 1.021$\pm$0.003 & 2.14 & & & &\\
LHS 2924 & C05,~R09 & M9.5 & Field & 1.016$\pm$0.004 & 2.23 & & & &\\
Gl 752B & C05,~R09 & M8 & Field & 1.022$\pm$0.004 & 2.14 & & & &\\
LP 412-31 & R09 & M8 & Field & 1.022$\pm$0.003 & 2.14 & & & &\\
\hline
Teide 1 & B10 &  M8 & 120 & 1.057$\pm$0.015 & 2.17 & 1.088 & 0.044\\
HII 1348B & G12 & M8 $\pm$ 1 & 120 & 1.068$\pm$0.023 & 2.19 & &\\ 
BPL~62 & P00, P03 & M8.3 & 120 & 1.138$\pm$0.009 & 2.15 & &\\
\hline
GSC08047 B & P12 & M9.5 & 30 & 1.004$\pm$0.006 & 2.18 & 1.004 & \\
\hline
USco~J161047$-$223949 & L08 & M9 & 11 & 1.012$\pm$0.007 & 2.16 & 1.005 & 0.014\\  
USco~J160648$-$223040 & L08 & M8 & 11 & 1.006$\pm$0.006 & 2.18 & & \\
USco~J160606$-$233513 & L08 & L0 & 11 & 1.018$\pm$0.006 & 2.17 & & \\
USco~J160714$-$232101 & L08 & L0 & 11 & 0.976$\pm$0.011 & 2.24 & & \\
USco~J160737$-$224247 & L08 & L0 & 11 & 0.988$\pm$0.007 & 2.22 & & \\
USco~J160818$-$223225 & L08 & L0 & 11 & 0.999$\pm$0.006 & 2.25 & & \\
USco~J160830$-$233511 & L08 & M9 & 11 & 1.024$\pm$0.004 & 2.17 & & \\
USco~J160847$-$223547 & L08 & M9 & 11 & 1.011$\pm$0.005 & 2.17 & & \\
USco~J161227$-$215640 & L08 & L0 & 11 & 0.999$\pm$0.019 & 2.22 & & \\
USco~J161302$-$212428 & L08 & L0 & 11 & 1.001$\pm$0.009 & 2.20 & & \\
USco~J155419$-$213543 & L08 & M8 & 11 & 1.015$\pm$0.005 & 2.17 & & \\
DENIS~161103$-$242642 & L08 & M9 & 11 & 0.996$\pm$0.006 & 2.19 & & \\
SCH~162528$-$165850 & L08 & M8 & 11 & 1.020$\pm$0.006 & 2.17 & & & \\
\hline\\
TWA 5B & N09 & M8/M8.5 & 8 & 1.027$\pm$0.004 & 2.17 & 0.997 & 0.025 & \\ 
DENIS J124514.1-442907 & L07 & M9.5 & 8 & 0.974$\pm$0.008 & 2.25 &  & \\
2MASS 1139-3159 & L07 & M9 & 8 & 0.980$\pm$0.003 & 2.23 &\\
2M1207A & P12 & M8 & 8 & 1.008$\pm$0.003 & 2.23 &\\
\hline\\
CT Cha B & S08 & $\geq$M8 & 2 & 0.982$\pm$0.007 & 2.29 & 0.986 & 0.010\\
04574903 & L04b & M9.25 & 1-2 & 0.992$\pm$0.004 & 2.23 & & \\ 
KPNO-Tau 6$^{1}$ & L04b & M8.5 & 1-2 & 0.996$\pm$0.004 & 2.24 & & \\
KPNO-Tau 7$^{1}$ & L04b & M8.25 & 1-2 & 0.998$\pm$0.004 & 2.20 & & \\
KPNO-Tau 9 & L04b & M8.5 & 1-2 & 0.975$\pm$0.008 & 2.25 & &\\ 
LRL 405 & M07 & M8 & 1-2 & 0.975$\pm$0.005 & 2.25 & &\\
\hline\\
152$-$717 & This work & M9.5 $\pm$0.5 & 1 & 0.914$\pm$0.013 & 2.27 & 0.959 & 0.041\\
057-247 & L06 & $\geq$M9.5 & 1 & 0.918$\pm$0.038 & 2.22 & &\\
107-453 & L06 & M8.0$\pm$2.0 & 1 & 0.972$\pm$0.022 & 2.22 & &\\
137-532 & L06 & $>$M9.5 & 1 & 0.984$\pm$0.024 & 2.22 & &\\
183-729 & L06 & $\geq$M9.5 & 1 & 1.006$\pm$0.036 & 2.22 & &\\
\hline\\
$^{1}$ \scriptsize Possible IR excess\\
\end{tabular}
\end{minipage}
\label{tab:6}
\end{table*}

Among the Pleiades objects described in B10, Calar 3 has an exceptionally low H$_2$(K) index (0.978$\pm$0.020). This result may be associated with unresolved questions of Calar 3's binarity and variability (P03, ~\citet{bailer-jones01}). Therefore, we have excluded Calar 3 from the extended dataset.

Among the Upper Sco objects, USco~J154722$-$213914 has a relatively high H$_{2}$(K) index (1.041), typical of an older object. This object has a flux excess from 1.9$\mu$m $\rightarrow$ 2.1$\mu$m which L08 were unable to explain. This flux excess may have affected the object's H$_{2}$(K) index. Therefore, we have excluded USco~J154722$-$213914 from the extended dataset. Otherwise, the Upper Sco objects' H$_{2}$(K) indices are scattered about 1.00, reflected in a mean H$_{2}$(K) index of 1.005.

While some of the $\sim$8 Myr old TWA objects have relatively low H$_{2}$(K) indices, a number of Upper Sco objects have similarly low indices. The TWA objects are one of the smaller subsets in our extended dataset, and we would need more objects to obtain a more rigorous mean value of H$_{2}$(K) for this association. Note that 2M1207A is known to have a circumstellar disc~(\citet{gizis02};~\citet{sterzik04};~\citet{scholz05}).

Among the 1-2~Myr old objects, KPNO-Tau~6 and KPNO-Tau~7 have the largest H$_{2}$(K) indices. These objects (as well as KPNO-Tau~12) have strong H$\alpha$ emission, indicating the presence of an accretion disc~\citep{muzerolle05}. In addition, KPNO-Tau~6 and KPNO-Tau~7 have significant mid-IR excesses in the {\it Spitzer}/IRAC passbands~\citep{guieu07}. The youth of these objects is unquestioned. It is possible that their IR flux excesses slightly modify the shape of their K~band spectra, producing larger than expected H$_{2}$(K) indices, even though these objects form part of a sample with low extinction and no obviously unusual infrared spectroscopic features (K.Luhman, private comm.) That being so, the scatter in the H$_{2}$(K) index for the Taurus objects is very small. 

There is significant research to show that the ONC is very young, with an average age of $\leq$1 Myr (\citealt{prosser94};~\citealt{hillenbrand97};~\citealt{palla99};~\citealt{riddick07}). In general, the H$_{2}$(K) indices of the ONC PMOs in the extended dataset are as low, or lower, than those of the 1-2~Myr objects in the same dataset, indicating that they have similar surface gravities. We noted earlier that the Lyon models predict similar gravities for brown dwarfs and PMOs of a given age, with little dependence on mass. This therefore indicates that the age of the ONC PMOs is $\sim$1~Myr, on average.

\section{Discussion}
\subsection{The Brown Dwarf Calibrators}
\subsubsection{The Field Dwarfs}
BRI~0021's NaI pEW is substantially weaker than the same feature in the spectrum of the other field dwarf, 2MASS~0345. BRI~0021's rotational velocity (v sin i) is $\sim$ twice that of 2MASS~0345 (\citet{antonova08};~\citet{reid99}), and the strengths of stellar absorption lines are known to vary according to a star's rotational velocity and angle of inclination~\citep{guthrie65}, but the variation is unlikely to be sufficient to explain this discrepancy~\citep{stoeckley68}. Rapid rotation may alter the cloud structure, and thus may affect the observed spectrum. However, BRI~0021's H$_{2}$(K) index is similar to that of 2MASS~0345. We therefore conclude that both field dwarfs have similar surface gravities.

The ages of these objects are unknown but assuming an age of 1~Gyr for both field dwarfs, at a distance of $\sim$27pc~\citep{faherty09} and m$_{\rm{K}}$=12.7 for 2MASS~0345, and at a distance of $\sim$11.8pc~\citep{faherty09} and m$_{\rm{K}}$=10.6 for BRI~0021, the Lyon group DUSTY model isochrone predicts 0.08M$_{\odot}<$~M~$<$~0.09M$_{\odot}$ and 5.28~$<$~log g~$<$~5.32 for both objects. If these objects are as old as 10~Gyr, the mass range for both objects is unchanged, while their surface gravities now range from 5.29~$<$~log g~$<$~5.34.

\subsubsection{$\sigma$~Orionis Objects}
$\sigma$~Orionis~71 and $\sigma$~Orionis~J053$-$024 have flux excesses between 3.6$\mu$m and 8$\mu$m~\citep{caballero07}. These are strong indicators of the presence of a disc. Emission from a disc could weaken an object's NaI EW, thereby reducing its apparent surface gravity. If the strength of the EWs are due to gravity plus veiling, this effect should correlate inversely with age. However, the effect could introduce more scatter both through geometric (viewing angle) effects, and if disc dispersal timescales vary from cluster to cluster. (For the effect of dust emission on the H$_{2}$(K) index, see Section 6.3.1.) Since discs may persist for 10~Myr after birth (\citealt{jayawardhana99};~\citealt{sterzik04}), their detection is not inconsistent with these objects being categorised as the ``older" of the  young objects in this sample. 

$\sigma$~Orionis~51's NaI pEW and H$_{2}$(K) index are markedly different from the corresponding values for $\sigma$~Orionis~71 and $\sigma$~Orionis~J053$-$024. If $\sigma$~Orionis~51 is younger than its siblings, it is more likely to be a member of the $\sigma$~Orionis cluster, while admitting the likelihood that $\sigma$~Orionis~71 and $\sigma$~Orionis~J053$-$024 are also cluster members. The age of the $\sigma$~Orionis cluster is believed to be $\leq$ 10~Myr~\citep{barrado01}. If star formation occurred episodically over several Myr, it is not unreasonable to find objects with ages less than the maximum age of the cluster. Thus age spreads of a few Myr are plausible.

At a distance of $\sim$385pc~\citep{bejar11}, assuming an age of 3~Myr, and with m$_{\rm{K}}$=16.1 for $\sigma$~Orionis~51 and $\sigma$~Orionis~71, and 16.2 for $\sigma$~Orionis~J053$-$024, the Lyon group DUSTY model isochrones predict 0.017M$_{\odot}<$~M~$<$~0.018M$_{\odot}$. Assuming an age of 7~Myr, the model masses are 0.022M$_{\odot}<$~M~$<$~0.024M$_{\odot}$. Over this age range, the model gravities are 3.75~$<$~log g~$<$~3.93. 

\subsubsection{The 1-2 Myr Objects}
2MASS~0535$-$0546, a member of the ONC, is an eclipsing binary in which the masses and radii of the components have been determined~\citep{stassun06}. Thus the surface gravity of each component can be precisely known. For both primary (A1) and secondary (A2), the surface gravity is consistent with youthful objects, log g (A1)~$=$~3.52~$\pm$~0.03, log g (A2)~$=$~3.54~$\pm$~0.03 (\citealt{gomez09}, hereafter G09). Also, the large radii of the primary and secondary components (R1~$=$~0.675R$_{\rm{\odot}}$ $\pm$~0.023, R2~$=$~0.486R$_{\rm{\odot}}$ $\pm$~0.018 (G09)), are consistent with model predictions for 1~Myr old objects. 

The NaI pEW determined here also supports the case for 2MASS~0535$-$0546 being among the youngest objects in our sample. 2MASS~0535$-$0546 has an earlier spectral type than the rest of the sample (M7), implying that the binary has a higher T$_{eff}$. The K band NaI doublet is temperature sensitive in that it usually becomes much weaker in early L dwarfs (C05) and somewhat weaker in early M dwarfs but there does not appear to be a strong temperature dependence between M7 and M9 in that work. Therefore, it is reasonable to include the NaI pEW of this object with those of the other 1-2 Myr objects. We note that the doublet remains strong in our spectrum of the L0 dwarf 2MASS~0345 (see Figure 1 and Table 2).

KPNO-Tau~4 and KPNO-Tau~12, have NaI pEWs consistent with low surface gravity. Although KPNO-Tau~1 has a NaI pEW suggesting it has a higher surface gravity, its H$_{2}$(K) index is consistent with it being a low surface gravity object. This disagreement shows the danger of using single indicators to infer an age or even a surface gravity for individual objects.

The Taurus objects we chose for our sample, as well as those in the extended dataset, have a K band magnitude distribution that is typical for the objects in this cluster (\citealt{jayawardhana03}; L04b). Therefore we are confident we have not introduced a bias in favour of younger objects in selecting our sample.

At a distance of 140pc~\citep{guieu06}, with m$_{\rm{K}}$ between 13.7 and 14.9, for 1~Myr old objects the Lyon DUSTY model isochrones predict 0.02M$_{\odot}<$~M~$<$~0.03M$_{\odot}$ and 3.72 $<$~log g~$<$~3.79 for KPNO-Tau~1 and KPNO-Tau~4, and 0.006M$_{\odot}<$~M~$<$~0.007M$_{\odot}$ and 3.50~$<$~log g~$<$~3.51 for KPNO-Tau~12. At a distance of 435pc~\citep{stassun06}, with  m$_{\rm{K}}$ $ = $ 13.8, the Lyon DUSTY model isochrones predict a total mass M $ = $ 0.07M$_{\odot}$ and log g~$=$~3.54 for 2MASS~0535$-$0546. Note that the surface gravities have been calculated for both components of this binary and are in close agreement with the model values.
\\
\subsection{152$-$717}

\begin{figure}
\centering
\includegraphics[scale=0.36]{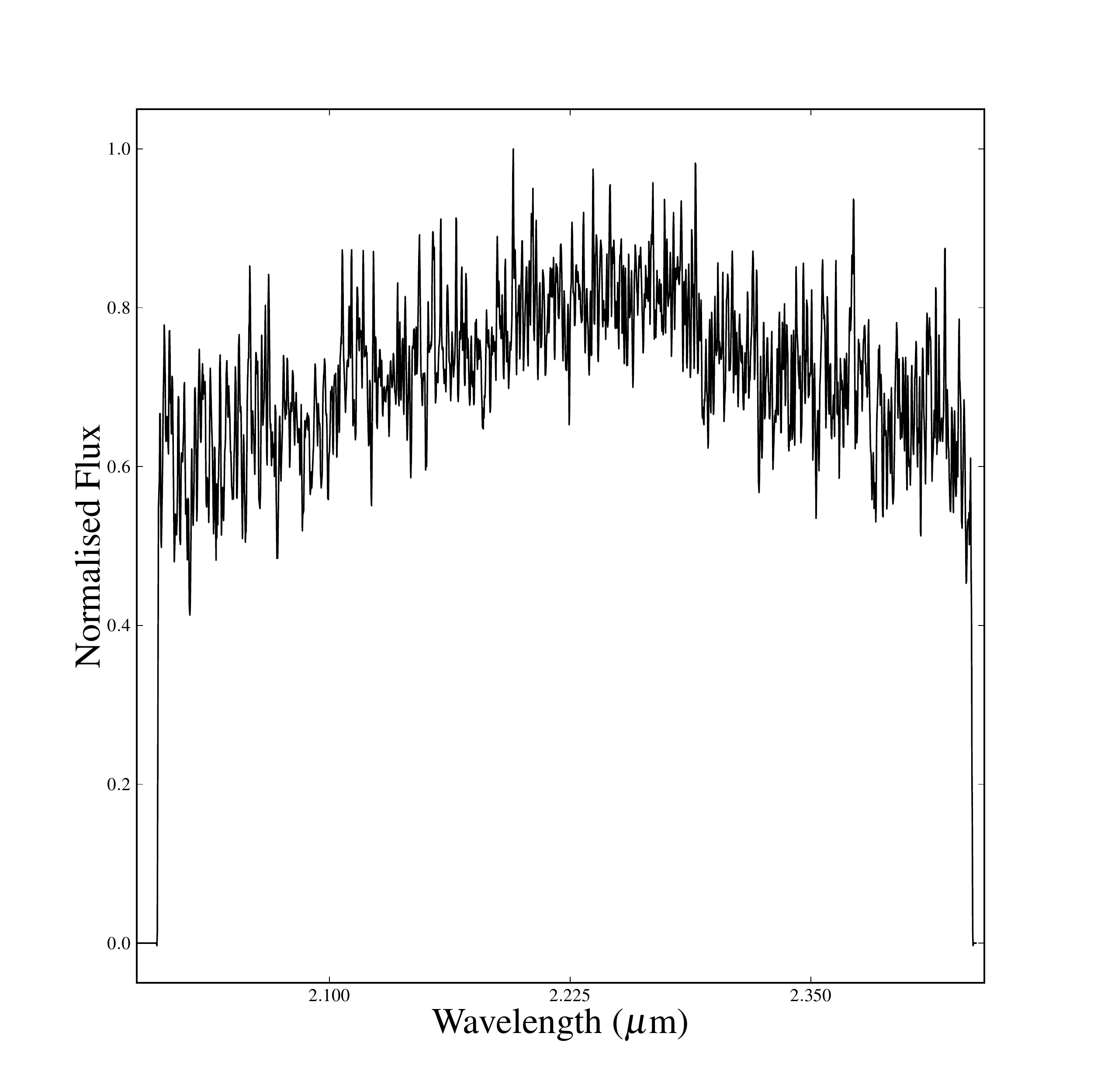} 
\caption{The spectrum of 152$-$717 after 3 pixel boxcar smoothing. With a S/N of 5, narrow, gravity-sensitive spectral features are not discernible.}
\label{fig:fig12}
\end{figure}

\begin{figure*}
\centering
\includegraphics[scale=0.45]{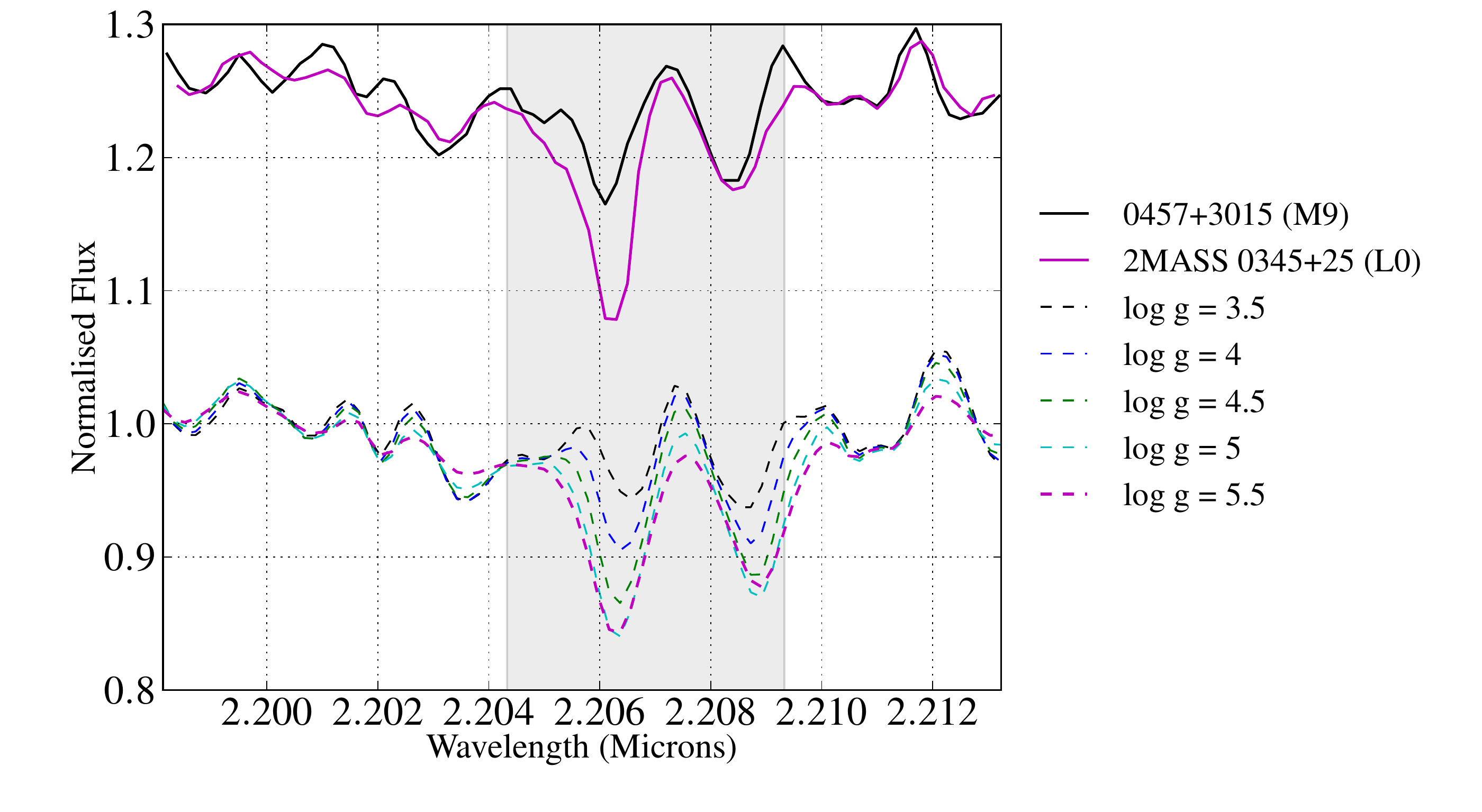}
\caption{Model spectra from S12 of the H$_2$O absorption lines adjacent to the NaI doublet (shaded region) compared to observations of a 1-2~Myr object (0457+3015) and field dwarf (2MASS~0345). Our object spectra have been offset from the model spectra to aid comparison.}
\label{fig:fig13}
\end{figure*}

Owing to the difficulty in maintaining a guide star lock even in the fainter parts of the Orion nebula, the integration time for 152$-$717 was shorter than requested. As a result, the extracted spectrum is not of sufficient quality to detect weak features such as the NaI doublet. Nonetheless, the spectrum shows the characteristic 
H$_2$O absorption of a young, late M-type object and the CO ($v$=2$-$0) absorption trough and the bandheads at 2.29$\mu$m, 2.32$\mu$m, 2.35$\mu$m and 2.38$\mu$m are clearly measured for the first time. See Figure 12.

152$-$717's H$_2$(K) index is at the lower end of the range of H$_2$(K) indices for the ONC PMOs in the extended dataset (Table 6). This is consistent with 152$-$717 being among the youngest of the ONC PMOs. However, as we only have a single surface gravity/age indicator, we caution against interpreting this object's H$_2$(K) index as proof of its youth. 

We were also able to derive a revised spectral type. We calculate that 152$-$717 has a spectral type M9.5$\pm1.0$. At a distance of $\sim$450pc~\citep{luhman00}, assuming an age of 1~Myr and  m$_{\rm{K}}$=17.6, the Lyon group DUSTY model isochrone predicts 0.005M$_{\odot} <$ M $<$ 0.006M$_{\odot}$ and 3.47 $<$ log g $<$ 3.50 for this object (\citealt{chabrier00};~\citealt{baraffe02}).

\subsection{Water Absorption in the K~band}
It was noted that a number of M-dwarf spectra exhibit similar spectral structure in the vicinity of the 2.206~$\mu$m NaI absorption feature. The pattern is a feature of late-type dwarfs. (Synthetic spectra produced for objects with  T$_{\rm{eff}}$ from 2500K$\;\rightarrow\;$3300K show this pattern weakening with increasing T$_{\rm{eff}}$~\citep{lyubchik12}). It is reproduced in the synthetic spectra used to model CIA opacity in this paper [S12] and arises from water absorption (D.Saumon, private communication). See Figure 13.

\section{Comparing the Age Indicators}
We used our dataset of calibrator brown dwarfs to study the merits of each age indicator. We computed the mean value of NaI pEW and H$_{2}$(K) index for the Taurus and $\sigma$~Orionis objects and found that the means agreed to within 0.75$\sigma$. This increased to 1.31$\sigma$ for the H$_{2}$(K) index.  While the H$_{2}$(K) index performs significantly better, neither method categorically differentiates the populations.

When we carried out the same exercise for the Taurus objects and our field dwarfs, the confidence level rose to 1.86$\sigma$ (NaI pEW) and 6.76$\sigma$ (H$_{2}$(K) index).

The dataset of brown dwarf calibrators contains only a few objects in each age group.  To minimise the uncertainty in the H$_{2}$(K) index, we examined the larger population of objects contained in our extended dataset. We compared the mean H$_{2}$(K) index for the 1-2 Myr objects in the extended dataset with that of the Upper Sco objects and derived a 3.37$\sigma$ confidence that these objects are from separate populations. The confidence level increased to 3.96$\sigma$ when we compared the means of the 1-2 Myr objects with those of the Pleiades objects in our sample. The Pleiades have a wide scatter due to one object. This scatter could be reduced with the addition of more high quality K band spectra.

These results suggest that the H$_{2}$(K) index is more gravity-sensitive than the NaI pEW. We have shown that the mean H$_{2}$(K) index can change significantly between 1-2 Myr clusters and clusters $\geq$10 Myrs. It therefore appears to be more gravity-sensitive at very young ages than the triangular H-band continuum which is found in cluster objects with ages $\leq$ 5 Myr and in young field dwarfs, but which can persist for several tens of Myr (A07). The H$_{2}$(K) index is applicable to lower resolution spectra. This has allowed us to suggest that the PMO in our original dataset has low surface gravity. Since narrow spectral features were not resolved in the spectrum of this PMO, this would not have been possible had we depended only on the NaI pEW of the object.

We conclude that the H$_{2}$(K) index is a more sensitive age indicator.

\section{Conclusions}
The K~band slope is a good indicator of an object's surface gravity, provided the object has solar metallicity. While most nearby star formation regions tend to have solar metallicity, individual objects' metallicities should be taken into account when considering the object's surface gravity. 

We have defined a new spectral index, H$_{2}$(K), which is a good indicator of surface gravity, particularly for the youngest sources (1-10~Myr), where it is more sensitive to surface gravity than the triangular H~band peak. It is at least as good as the equivalent width of the 2.21$\mu$m NaI doublet in differentiating the surface gravities of late-type objects, and does so with less scatter and without the need for high-resolution spectra. This has allowed us to examine other datasets, obtaining similar results to those obtained from our own brown dwarf calibrators. All datasets show some scattering of H$_{2}$(K) indices within a cluster. Considering the uncertainties associated with the data, it is unclear whether or not this scatter reflects a genuine spread of ages.

We have shown that the H$_{2}$(K) index and the EW of the 2.21$\mu$m NaI doublet can find the difference, at least statistically, between a population of $\sim$1~Myr objects and field dwarfs. In addition, the H$_{2}$(K) index can statistically differentiate a population of $\sim$1~Myr objects and a population of $\sim$10~Myr objects. The H$_{2}$(K) index can also be used to separate low-mass members from foreground and background objects in young clusters and associations.

By comparing the H$_{2}$(K) indices of 152$-$717 and the ONC PMOs from L06 to the H$_{2}$(K) indices of our 1-2~Myr calibrators, we infer that, on average, the ONC PMOs and our 1-2~Myr calibrators have a similar age.

The $\sigma$~Orionis objects' H$_{2}$(K) indices and NaI pEWs are consistent with one object being significantly younger than the others. If true, this would indicate a variation of ages within the cluster, rather than suggesting that the older objects are not cluster members. Indeed, if these objects were contaminants, they could only be much older field objects, and this is clearly not the case.

Given the differences in H$_{2}$(K) index among the Pleiades objects and that some of their spectra have quite low S/N in the K band, it would be worthwhile obtaining higher S/N K band spectra of objects in clusters with well constrained ages so that these surface gravity indicators can be investigated. 

We have shown that it is not necessary to have high-resolution spectra of these objects to be able to constrain the size of their population in young clusters and the low-mass end of the IMF.

We now have two gravity-sensitive methods of differentiating late-type dwarfs, the EW of neutral alkali metal lines and the H$_{2}$(K) index, both of which show good correlations with age and which will allow researchers to ascribe statistical ages to samples of brown dwarfs in pre-main sequence clusters. These can then be used to distinguish between planetary mass objects and somewhat older brown dwarfs with a similar luminosity and temperature. 

An object of future research is to determine the extent to which scatter in these measures of surface gravity is due to an age spread.

\section*{Acknowledgements}
We thank the referee for their helpful comments and suggestions.

This paper is based on observations obtained in programmes GN-2008B-Q-20 and GN-2009B-Q-57 at the Gemini Observatory, which is operated by the Association of Universities for Research in Astronomy Inc., under a cooperative agreement with the NSF on behalf of the Gemini partnership: The National Science Foundation (USA), the Science and Technology Facilities Council (UK), the National Research Council (Canada), CONICYT (Chile), the Australian Research Council (Australia). CNPq (Brazil) and CONICET (Argentina).

J. I. Canty is supported by a University of Hertfordshire PhD studentship.

\bibliographystyle{mn2e}

\label{lastpage}

\end{document}